\documentclass[a4paper]{article}
\usepackage{amsmath}
\usepackage{amstext}
\usepackage{amssymb}
\usepackage{graphicx}

\begin{document}

\title{Probing dynamical spacetimes\\ with\\ gravitational waves\footnote{Chapter to appear in the Springer Handbook of Spacetime, Springer Verlag, 2013.}\\}
\author{Chris Van Den Broeck\footnote{\texttt{vdbroeck@nikhef.nl}}\\ \\
Nikhef -- National Institute for Subatomic Physics,\\ Science Park 105, 1098 XG Amsterdam, The Netherlands}

\maketitle

\begin{abstract}
This decade will see the first direct detections of gravitational waves by observatories such as Advanced LIGO and Virgo. Among the prime sources are coalescences of binary neutron stars and black holes, which are ideal probes of dynamical spacetime. This will herald a new era in the empirical study of gravitation. For the first time, we will have access to the genuinely strong-field dynamics, where low-energy imprints of quantum gravity may well show up. In addition, we will be able to search for effects which might only make their presence known at large distance scales, such as the ones that gravitational waves must traverse in going from source to observer. Finally, coalescing binaries can be used as cosmic distance markers, to study the large-scale structure and evolution of the Universe.

With the advanced detector era fast approaching, concrete data analysis algorithms are being developed to look for deviations from general relativity in signals from coalescing binaries, taking into account the noisy detector output as well as the expectation that most sources will be near the threshold of detectability. Similarly, several practical methods have been proposed to use them for cosmology. We explain the state of the art, including the obstacles that still need to be overcome in order to make optimal use of the signals that will be detected. Although the emphasis will be on second-generation observatories, we will also discuss some of the science that could be done with future third-generation ground-based facilities such as Einstein Telescope, as well as with space-based detectors.
\end{abstract}

\section{Introduction}

General relativity (GR) is a highly non-linear, dynamical theory of gravitation. Yet, until the 1970s, almost all tests of GR were theoretically based on the behavior of test particles in a \emph{static} gravitational field \cite{MTW}. These include the perihelium precession of Mercury, deflection of star light by the Sun, and Shapiro time delay. The parameterized post-Newtonian (PPN) formalism (for an overview, see \cite{Will2006}) was developed to provide a systematic framework for these and other checks, by appropriately parameterizing various aspects of spacetime geometry viewed as a non-linear superposition of contributions from \emph{e.g.}~the Sun and the planets. Even so, the most important early experiments did not require much more than an expansion of the Schwarzschild metric in $GM/(c^2 r)$ up to the first few sub-leading terms, with $M$ the mass and $r$ the distance. Although excellent agreement with theory was obtained, the aspects of GR that were actually tested were somewhat limited, mostly amounting to the influence on the motion of test particles of low-order general-relativistic corrections to the Newtonian gravitational field.

The situation improved with the discovery of the Hulse-Taylor binary neutron star\index{Hulse-Taylor binary neutron star} in 1974 \cite{Taylor1982}. One of the two stars can be observed electromagnetically as a pulsar, and from this signal it was inferred that the orbital motion of the binary changes as predicted by GR, assuming that orbital energy and angular momentum are being carried away by gravitational waves (GW). This was an event of historic significance, as it provided incontrovertible evidence of the dynamical nature of the gravitational field. Subsequently, similar and even more relativistic binary neutron stars were discovered, allowing for new tests of GR in a ``post-Keplerian" framework \cite{Freire2012}. Nevertheless, explaining the observed dissipative dynamics related to gravitational wave emission only requires a lowest-order approximation to GR in powers of $v/c$, with $v$ a characteristic velocity.

Some of the most exciting aspects of general relativity still remain out of our empirical reach. What we would like to explore is the full non-linear dynamics of the gravitational field itself, including its self-interaction. From this perspective, even the most relativistic binary neutron star system that is currently known, PSR J0737-3039\index{PSR J0737-3039} \cite{Freire2012}, is still in the relatively slowly varying, weak-field regime, with an orbital compactness of $GM/(c^2 R) \simeq 4.4 \times 10^{-6}$ (where $M$ is the total mass and $R$ the orbital separation), and a typical orbital speed $v/c \simeq 2 \times 10^{-3}$.\footnote{For comparison, the surface gravity of the Sun is $\simeq 2 \times 10^{-6}$, and the orbital speed of the Earth as it moves around the Sun is $v/c \simeq 10^{-4}$.} By contrast, for a compact binary just prior to the final plunge and merger one has $GM/(c^2 R)  \sim 0.2$ and $v/c \sim 0.4$. This will bring us to the genuinely strong-field, dynamical regime of general relativity, which in the foreseeable future will only be accessible by means of gravitational wave detectors.

The first detection of gravitational waves by the Advanced LIGO\index{Advanced LIGO} and Virgo\index{Advanced Virgo} detectors might happen as early as 2015, and certainly before the end of the decade \cite{ratespaper}. Around 2020, a network of five large interferometric GW detectors will be in place: in addition to the two Advanced LIGO interferometers \cite{LIGO} and Advanced Virgo \cite{Virgo}, there will be the Japanese KAGRA\index{KAGRA} \cite{KAGRA}, and IndIGO\index{IndIGO} in India \cite{IndIGO}. There is also the smaller GEO-HF\index{GEO-HF} in Germany \cite{GEO}. These are usually referred to as second-generation detectors. A conceptual design study for a third-generation observatory called Einstein Telescope\index{Einstein Telescope} (ET) was recently concluded \cite{ET}, and there are plans for a space-based observatory called LISA\index{LISA} \cite{LISA}. There is a considerable body of literature on the projected capabilities of ET and LISA in probing the dynamics of gravity. Although attention will be given to these, our main focus in this chapter will be on what can be achieved with the upcoming advanced detectors. In particular, in the last few years, development has started of hands-on data analysis techniques for use on signals detected with second-generation observatories, properly taking into account the noisy detector output as well as the expectation that most sources will be near the threshold of detectability.

Since the advent of GR, a large number of alternative theories of gravity\index{alternative theories of gravity} have been proposed; for a partial list, see \cite{Yunes2009}. Among these, GR tends to be the simplest and the most elegant; moreover, many of the alternatives are already strongly constrained by existing observations. Consequently, our primary aim will not be to seek confirmation of an alternative theory and measure its parameters; rather, we want to develop a test of GR itself. The testing should be as generic as we can make it, in the sense that if macroscopic deviations from GR exist, we want to find them even if they take a form that is yet to be envisaged, rather than looking inside a class of particular, existing alternative theories. That said, we will occasionaly mention the predictions of such alternative models to indicate the power of the probe that direct gravitational wave detection will provide us with. 

Recently proposed tests of the strong-field dynamics broadly fall into two categories. One consists of checking for possible alternative polarization states\index{polarization states} beyond the two polarizations that GR predicts, and which might only make their presence known in the case of gravitational waves that were generated in the ultra-relativistic regime. The other category focuses on the coalescence process of compact binary objects (neutron stars and black holes) \cite{Will2006}.  

Searching for alternative polarizations started in earnest with the detailed studies made on the electromagnetically observed binary neutron stars. Here we will explain how one would go about looking for their signature in data from gravitational wave detectors, in particular using transient signals such as will be emitted by supernova explosions or, again, coalescing compact binaries. There have also been studies about how to use \emph{stochastic} gravitational waves for this purpose \cite{Nishizawa2009}; these could take the form of a primordial GW background, or they could be a ``bath" of radiation caused by a large number of unresolvable astrophysical sources, such as the combined population of all compact binary coalescence\index{compact binary coalescence} events, or cosmic string cusps. Due to space limitations, here we will limit ourselves to resolvable transient sources. Although with a single interferometric detector one would not be able to tell the difference between, or even measure, additional polarizations, this does become possible with a \emph{network} of detectors. In particular, it is possible to combine the outputs of multiple interferometers to construct a so-called \emph{null stream},\index{null stream} which should be devoid of signals if the only polarization states\index{polarization states} present are the ones predicted by GR. More generally, one can have null streams which in addition to the usual tensor polarizations also exclude one or more of the alternative ones, allowing one to tell them apart. Here we will partially follow the recent discussions by Chatziioannou, Yunes, and Cornish \cite{Chatziioannou2012}, and by Hayama and Nishizawa \cite{Hayama2012}.

The coalescence of compact binaries consists of three regimes: an adiabatic \emph{inspiral}, the \emph{merger} leading to the formation of a single black hole (or an exotic alternative object!), and the \emph{ringdown} of the resulting object as it evolves towards a quiescent state. The inspiral regime is reasonably well understood thanks to the so-called post-Newtonian formalism\index{post-Newtonian formalism}  \cite{Blanchet2006}, in which physical quantities such as energy and flux are expanded in powers of $v/c$. A test of GR could then take the form of identifying directly measurable quantities, such as post-Newtonian coefficients in an expansion of the orbital phase,\index{orbital phase} which in GR are inter-related, and checking whether the predicted relationships really hold. This would constitute a very generic test of GR, in which no recourse needs to be taken to any particular alternative theory of gravity. Such a test was first proposed by Arun \emph{et al.} \cite{Arun2006a}. Next, Yunes and Pretorius developed the \emph{parameterized post-Einsteinian} (ppE) framework,\index{parameterized post-Einsteinian framework} which considerably generalized the family of waveforms used in \cite{Arun2006a} by allowing for a larger class of parameterized deformations \cite{Yunes2009,Cornish2011}. The basic idea of Arun \emph{et al.} was implemented in a Bayesian way by Li \emph{et al.} using waveforms in the ppE family \cite{Li2012a,Li2012b}. The latter approach focuses on hypothesis testing. This has the advantage that since for every detected sources the same ``yes/no" question is asked, evidence for or against GR has a tendency to build up as information from an increasing number of detections is included. 

Moving beyond the inspiral regime, the ringdown can be studied in a variety of ways. In particular, the \emph{no hair theorem} can be tested, which says that in GR, the spacetime around a quiescent, electrically neutral black hole is determined uniquely by its mass and spin \cite{Hansen1974,Ryan1997}. The ringdown process can be modeled as perturbations on a fixed black hole spacetime, and the Einstein field equations impose relationships between the ringdown frequencies and damping times of the various modes that can get excited \cite{Gossan2012}. Verifying these inter-dependences amounts to testing the no hair theorem. Moreover, it has been shown that the \emph{amplitudes} of the ringdown modes carry information about the masses and the spins of the binary compact object that merged to form a single black hole \cite{Kamaretsos2012b}. As shown earlier by Ryan, the no hair theorem can also be tested by monitoring the motion of a small object (a neutron star or a stellar mass black hole) around a very massive black hole or exotic object, effectively mapping out the latter's spacetime geometry \cite{Ryan1997}.
  
As demonstrated by Schutz already in 1986, inspiraling and merging compact binaries can also be used as cosmic distances markers, or \emph{standard sirens}, to probe the large-scale structure and evolution of the Universe \cite{Schutz1986}. A variety of techniques have been proposed to exploit this fact using the second-generation detectors, ET, and space-based detectors. The second-generation observatories will mainly give us information about the Hubble constant\index{Hubble constant} $H_0$; however, they will do so in a way that is completely independent of existing measurements, and in particular does not require the so-called cosmic distance ladder,\index{cosmic distance ladder} with its potentially unknown systematic errors at every rung \cite{Nissanke2010,Taylor2012a,DelPozzo2012}. In the case of ET and space-based detectors, it is also possible to study the matter content of the Universe \cite{Sathyaprakash2010,Zhao2011,Taylor2012b,Messenger2012a,MacLeod2008}. By now we know that the expansion of the Universe is accelerating \cite{Riess1998}, which can be modeled heuristically by invoking a new substance called \emph{dark energy}.\index{dark energy} An exciting prospect is probing the equation of state of dark energy with gravitational waves, again in a way that is independent of conventional observations. 

This chapter is structured as follows. In Sec.~\ref{sec:polarizations} we discuss how one might look for alternative polarization states in transient GW signals, using a network of detectors. Next, we explain how the inspiral of compact binaries can be used to arrive at a very generic test of the strong-field dynamics of general relativity, including self-interaction (Sec.~\ref{sec:self-interaction}). The emphasis will be on second-generation detectors, where most sources will be near the threshold of detectability. As we shall see, in the case of binary neutron stars, appropriate data analysis methods are already in place, which can be applied to raw data from the advanced detectors as soon as they become available. Binary black holes are dynamically far richer, but they also pose formidable data analysis problems. A discussion of merger and ringdown, and tests of the no hair theorem, will naturally bring us to third-generation ground-based observatories as well as space-based detectors (Sec.~\ref{sec:nohair}), and we will give an overview of what one might expect from them. In Sec.~\ref{sec:cosmography}, we will briefly recall the basics of modern cosmology, and investigate what gravitational wave observations might have to say about the evolution of the Universe. A summary and conclusions are given in Sec.~\ref{sec:summary}.

We will denote binary neutron stars\index{binary neutron stars} by BNS, neutron star-black hole systems\index{neutron star-black hole systems} by NSBH, and binary black holes\index{binary black holes} by BBH. The usual post-Newtonian notation will be employed, where ``$q$PN order" with $q$ integer or half-integer refers to $\mathcal{O}\left[(v/c)^{2q}\right]$ beyond leading order in expansions in $v/c$. Unless stated otherwise, we use units such that $G = c = 1$.

\section{Alternative polarization\index{polarization states} states}
\label{sec:polarizations}

In the so-called transverse-traceless gauge, the metric perturbation only has spatial components, and for a signal propagating in the $z$ direction in a coordinate system associated with unit vectors $(\hat{e}_x, \hat{e}_y, \hat{e}_z)$, it can be brought in the form \cite{MTW}
\begin{equation}
h^{\rm TT}_{ij} = h_+ {\rm e}^+_{ij} + h_\times {\rm e}^\times_{ij},
\end{equation}
with 
\begin{eqnarray}
{\rm e}^+_{ij} &=& \hat{e}_x \otimes \hat{e}_x - \hat{e}_y \otimes \hat{e}_y, \\
{\rm e}^\times_{ij} &=& \hat{e}_x \otimes \hat{e}_y + \hat{e}_y \otimes \hat{e}_x. 
\end{eqnarray}
$h_+$ and $h_\times$ are the magnitudes of the independent ``plus" and ``cross" tensor polarizations, respectively. The response to a gravitational wave of an L-shaped interferometric detector is a linear combination of these:
\begin{equation}
h(t) = F_+ h_+(t) + F_\times h_\times(t).
\end{equation}
The \emph{beam pattern functions}\index{beam pattern functions} $F_+$, $F_\times$ depend on the sky position $\hat{\Omega} = (\theta, \phi)$ of the source:\footnote{In addition, there is a rotational degree of freedom $\psi$ around the axis defined by $\hat{\Omega}$, but without loss of generality we can take $\psi = 0$.}
\begin{eqnarray}
F_+ &=& \frac{1}{2} (1+\cos^2\theta)\,\cos 2\phi, \\
F_\times &=& -\cos\theta \sin 2\phi. 
\end{eqnarray}

\begin{figure}[htbp!]
\centering
\includegraphics[height=6cm]{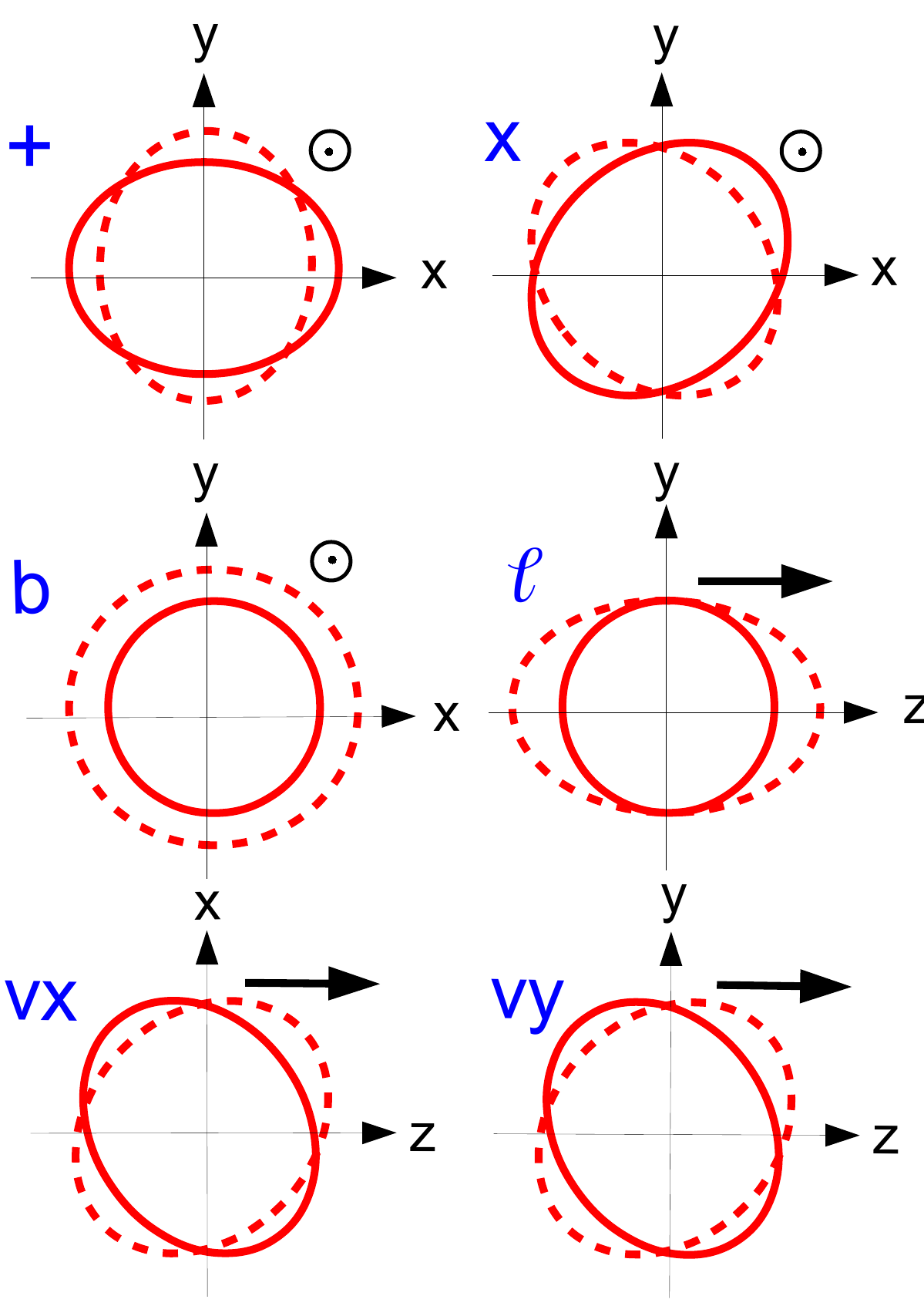}
\caption{In metric theories of gravity, up to 6 polarization states\index{polarization states} are allowed. At top left and right, we illustrate the transverse ``$+$" and ``$\times$" tensor polarizations. At middle left, the transverse ``breathing" mode is shown, and at middle right the longitudinal scalar mode. At bottom left and right, one has the vector modes. (Adapted from \cite{Will2006}.)}
\label{fig:Polarizations}
\end{figure}

In metric theories of gravity, up to 6 degrees of freedom are allowed \cite{Will2006}; these are illustrated in Fig.~\ref{fig:Polarizations}. Other than the plus and cross polarizations, they include a scalar ``breathing mode"\index{breathing mode} (``${\rm b}$"), a scalar longitudinal mode\index{longitudinal mode} (``$\ell$"), and vectorial modes\index{vectorial modes} (``${\rm vx}$", ``${\rm vy}$"). The full metric perturbation then takes the form (see \cite{Nishizawa2009} and references therein)
\begin{equation}
h_{ij} = h_+ {\rm e}^+_{ij} + h_\times {\rm e}^\times_{ij} + h_{\rm b} {\rm e}^{\rm b}_{ij} + h_{\ell} {\rm e}^{\ell}_{ij} + h_{\rm vx} {\rm e}^{\rm vx}_{ij}  + h_{\rm vy} {\rm e}^{\rm vy}_{ij},
\end{equation}  
with
\begin{eqnarray}
{\rm e}^{\rm b} &=& \hat{e}_x \otimes \hat{e}_x + \hat{e}_y \otimes \hat{e}_y, \\
{\rm e}^{\rm \ell} &=& \sqrt{2}\,\hat{e}_z \otimes \hat{e}_z, \\
{\rm e}^{\rm vx} &=& \hat{e}_x \otimes \hat{e}_z + \hat{e}_z \otimes \hat{e}_x, \\
{\rm e}^{\rm vy} &=& \hat{e}_y \otimes \hat{e}_z + \hat{e}_z \otimes \hat{e}_y.
\end{eqnarray}
The full response of an interferometer to a signal containing all these polarization states\index{polarization states} is 
\begin{equation}
h = F_+ h_+ +  F_\times h_\times + F_{\rm b} h_{\rm b} + F_{\ell} h_{\ell} + F_{\rm vx} h_{\rm vx} + F_{\rm vy} h_{\rm vy},
\label{full}
\end{equation}
and
\begin{eqnarray}
F_{\rm b} &=& -\frac{1}{2} \sin^2\theta\, \cos 2\phi, \label{bmode}\\
F_{\ell} &=& \frac{1}{\sqrt{2}} \sin^2 \theta\, \cos 2\phi,  \label{lmode}\\
F_{\rm vx} &=& - \frac{1}{2} \sin 2\theta\,\cos 2\phi, \\
F_{\rm vy} &=& \sin\theta\,\sin 2\phi. 
\label{beampatternfunctions}
\end{eqnarray}

Currently, observational constraints on additional polarization modes are limited. From the Hulse-Taylor double neutron star\index{Hulse-Taylor binary neutron star} we know that the energy loss due to non-tensor emission must be less than 1\% \cite{Will2006,Hayama2012}. However, alternative polarizations might show up in more weak-field regimes and after having propagated over distances much larger than the characteristic scale of the Hulse-Taylor binary. Alternatively, they might only appear in situations where the source is far more relativistic, with high characteristic velocities $v/c$. In core collapse supernovae, radial velocities $v/c \sim 0.25$ are attained \cite{Fryer2006}, which may excite longitudinal modes.\index{longitudinal mode} In the case of binary inspiral, $v/c > 0.4$ is reached before the final plunge, which (using Kepler's third law) corresponds to a gravitational wave frequency $f = c^3 (v/c)^3/(\pi G M)$, with $M$ the total mass. For binary neutron stars\index{binary neutron stars} with component masses $(1.4,1.4)\,M_\odot$ this approximately equals 1600 Hz, which is in the sensitivity band of ground-based detectors. 

There are a number of alternative theories of gravity\index{alternative theories of gravity} which predict non-standard polarization states.\index{polarization states} To name but a few:
\begin{itemize}
\item Brans-Dicke theory\index{Brans-Dicke theory} is a scalar-tensor theory\index{scalar-tensor theories} of gravity which has scalar modes \cite{Maggiore2000}.
\item Scalar modes also occur in Kaluza-Klein theory,\index{Kaluza-Klein theory} where our 4D world arises after compactification of one or more extra spatial dimensions \cite{Alesci2005}.
\item Certain brane world models, such as the Dvali-Gabadadze-Porrati model\index{Dvali-Gabadadze-Porrati model} in the self-accelerating branch, have all six modes above \cite{Charmousis2006}.
\end{itemize}

A \emph{single} interferometric detector would not suffice to disentangle all these polarization states. To see this, consider a breathing mode\index{breathing mode} (see Fig.~\ref{fig:Polarizations}) impinging upon a detector, coming from a direction that is perpendicular to the plane of the interferometer. Then both detector arms would get lengthened and shortened in unison, but what an interferometer senses is the relative \emph{difference} in arm length. Or, consider a breathing mode\index{breathing mode} whose propagation direction corresponds to the orientation of one of the arms. Then this arm would be unaffected, while the other arm would still periodically get lenghtened and shortened, leading to a relative difference in arm lengths, which however would be indistinguishable from the effect of a gravitational wave with ``plus" polarization. Hence, a \emph{network} of interferometers is called for. 

Consider $D$ detectors at different positions on the Earth and whose noise is uncorrelated. A signal would reach the interferometers at different times. However, if one knew the sky position $\hat{\Omega}$, \emph{e.g.} because of an electromagnetic counterpart to the gravitational wave signal as might be expected from a conveniently oriented BNS or NSBH event \cite{Nakar2007}, then one would know how to time shift the outputs of the detectors to analyze the signal at a fixed time, say the arrival time at the Earth's center. Since from now on we assume $\hat{\Omega}$ to be known, we omit any explicit reference to it in expressions below. In each detector $A = 1, \ldots, D$, the output will take the form 
\begin{equation}
\bar{d}^A(f) = \bar{h}^A(f) + \bar{n}^A(f), 
\label{signal+noise}
\end{equation}
where each of the $\bar{h}_A$ takes the general form (\ref{full}), and the $\bar{n}_A$ represent the noise in each of the detectors. The overbar indicates that (a) we  will be considering the Fourier transforms of the relevant quantities, which are functions of frequency $f$ rather than time $t$, and (b) the data streams have been divided  by $\sqrt{S_A(f)}$, with $S_A(f)$ the \emph{noise spectral density}\index{noise spectral density} (basically the variance of the noise as a function of frequency) for detector $A$. The latter ensures that we will not have to worry about differences in design and operation between the various detectors. 

Eq.~(\ref{signal+noise}) can be expressed  in terms of the beam pattern functions:\index{beam pattern functions}  
\begin{equation}
\bar{d}^A(f) = \bar{F}^A_a h_a(f) + \bar{n}^A(f), 
\end{equation}
where $a = 1, \ldots, 6$ runs over the polarization states\index{polarization states}  ``$+$", ``$\times$", ``${\rm b}$", ``$\ell$", ``${\rm vx}$",  and ``${\rm vy}$", and summation over repeated indices is assumed. The first term in the right hand side expresses the signal as a linear combination of five vectors in the $D$-dimensional space of detector outputs: ${\bf \bar{F}}_+$, ${\bf \bar{F}}_\times$, ${\bf \bar{F}}_{\rm vx}$, ${\bf \bar{F}}_{\rm vy}$, and ${\bf \bar{F}}_{\ell}$; indeed, from Eqns.~(\ref{bmode}), (\ref{lmode}), it is clear that ${\bf \bar{F}}_{\ell} = - \sqrt{2}\,{\bf \bar{F}}_{\rm b}$, so that one only has one independent vector pertaining to the scalar modes. Also note that the remaining vectors can be linearly independent only if $D \geq 5$.

\begin{figure}[htbp!]
\centering
\includegraphics[height=5cm]{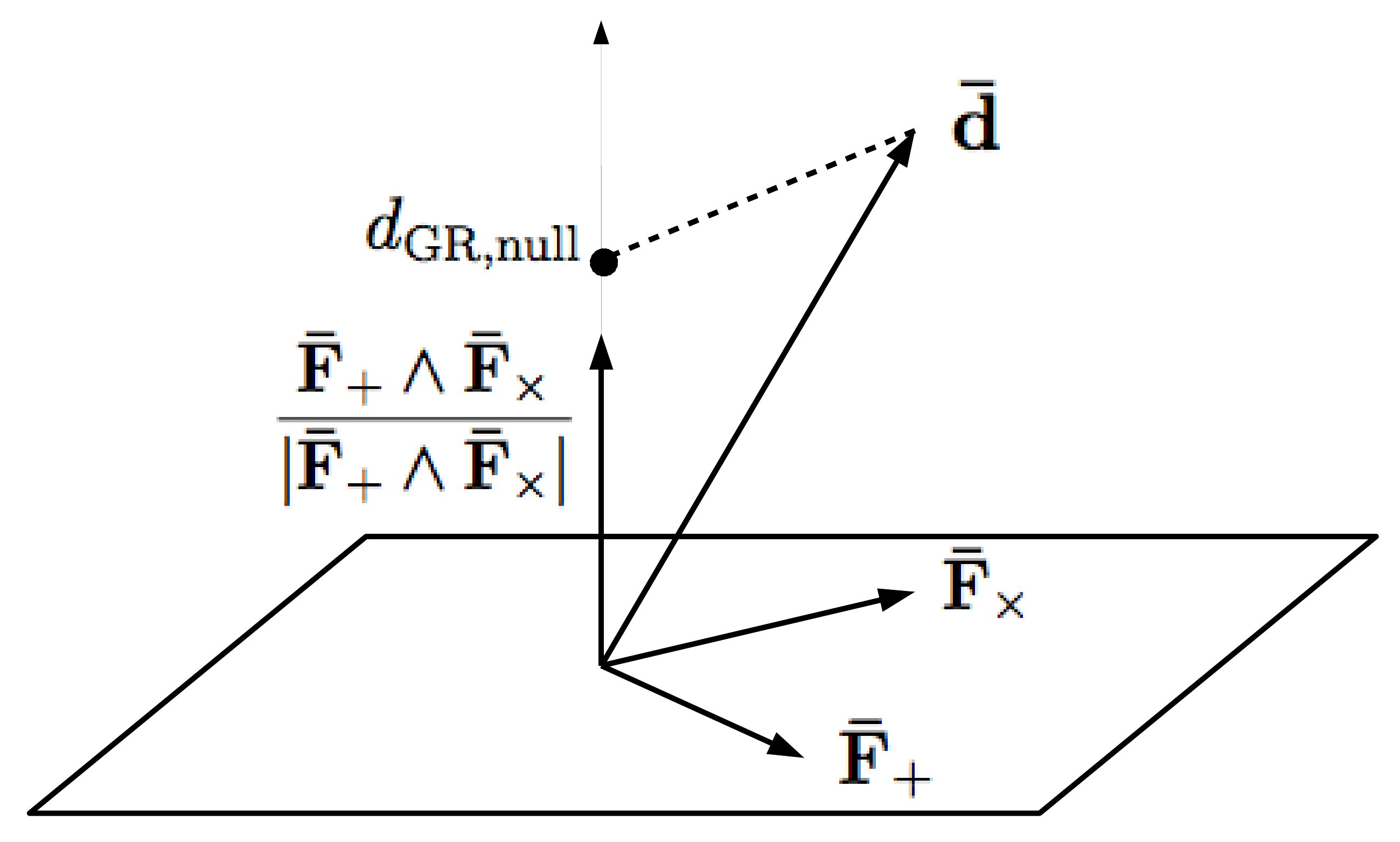}
\caption{An illustration of the construction of the null stream $d_{\rm GR,null}$ from a 3-detector output. The vector of outputs ${\bf \bar{d}}$, and the beam pattern vectors ${\bf \bar{F}}_+$ and ${\bf \bar{F}}_\times$, live in a 3-dimensional space. The null stream is obtained by projecting ${\bf \bar{d}}$ onto the unit normal to the plane determined by ${\bf \bar{F}}_+$, ${\bf \bar{F}}_\times$. The projection is guaranteed not to contain tensorial polarization modes.}
\label{fig:nullstream}
\end{figure}

In general relativity, only the tensor modes $h_+$ and $h_\times$ are present. Given three detectors (\emph{e.g.} the two Advanced LIGOs\index{Advanced LIGO} and Advanced Virgo,\index{Advanced Virgo} which will be the first to take data), a \emph{null stream}\index{null stream} can be constructed by eliminating these modes from the output vector ${\bf \bar{d}}$, following the original idea by G\"{u}rsel and Tinto \cite{Gursel1989}:
\begin{equation}
d_{\rm GR, null} = \frac{{\bf \bar{F}}_+ \wedge {\bf \bar{F}}_\times}{|{\bf \bar{F}}_+ \wedge {\bf\bar{F}}_\times|} \cdot {\bf \bar{d}},
\end{equation} 
where ${\bf \bar{F}}_+ \wedge {\bf \bar{F}}_\times$ is the vector whose components in the space of detector outputs are
\begin{equation}
\epsilon^{ABC} \bar{F}^B_+ \bar{F}^C_\times,
\label{wedgeproduct}
\end{equation}
with $\epsilon^{ABC}$ the completely antisymmetric symbol, and here too summation over repeated indices is understood. It is not difficult to see that $d_{\rm GR, null}$ can only contain non-standard polarizations; the tensor modes are projected out. This is illustrated in Fig.~\ref{fig:nullstream}. 
Hence, if GR is correct, $d_{\rm GR, null}$ should not contain a signal. If, on the other hand, one or more of the alternative polarizations $h_{\rm b}$, $h_{\ell}$, $h_{\rm vx}$, $h_{\rm vy}$ are present, then they will show up as a statistical excess in the null stream.\index{null stream}

Sometime after 2017, the Japanese KAGRA\index{KAGRA} will become active, and there will be four detectors, so that $D = 4$. This will allow for the construction of two null streams\index{null stream} which in addition to the tensor modes will also be devoid of \emph{e.g.} one of the two vector modes\index{vectorial modes} and one of the two scalar modes:  
\begin{eqnarray}
{}^{(4)}d^1_{\rm GR, null} &=& \frac{{\bf \bar{F}}_+ \wedge {\bf \bar{F}}_\times \wedge {\bf \bar{F}}_{\rm vx}}{|{\bf \bar{F}}_+ \wedge {\bf \bar{F}}_\times \wedge {\bf \bar{F}}_{\rm vx}|} \cdot {\bf \bar{d}}, \\
{}^{(4)}d^2_{\rm GR, null} &=& \frac{{\bf \bar{F}}_+ \wedge {\bf \bar{F}}_\times \wedge {\bf \bar{F}}_\ell}{|{\bf \bar{F}}_+ \wedge {\bf \bar{F}}_\times \wedge {\bf \bar{F}}_\ell|} \cdot {\bf \bar{d}}, 
\end{eqnarray}
where the wedge product is defined analogously to Eq.~(\ref{wedgeproduct}), but now using the 4-dimensional antisymmetric symbol $\epsilon^{ABCD}$. Note that for $D = 4$, there can not be a third independent null stream which also excludes the tensor modes.
 
Finally, around the end of the decade, IndIGO\index{IndIGO} may also be taking data, so that $D = 5$. In that case three null streams can be constructed that exclude the tensor modes: 
\begin{eqnarray}
{}^{(5)}d^1_{\rm GR, null} &=&  \frac{{\bf \bar{F}}_+ \wedge {\bf \bar{F}}_\times \wedge {\bf \bar{F}}_{\rm vx} \wedge  {\bf \bar{F}}_{\rm vy}}{|{\bf \bar{F}}_+ \wedge {\bf \bar{F}}_\times \wedge {\bf \bar{F}}_{\rm vx} \wedge  {\bf \bar{F}}_{\rm vy}|} \cdot {\bf \bar{d}}, \\
{}^{(5)}d^2_{\rm GR, null} &=& \frac{{\bf \bar{F}}_+ \wedge {\bf \bar{F}}_\times \wedge {\bf \bar{F}}_{\rm vx} \wedge {\bf \bar{F}}_\ell}{|{\bf \bar{F}}_+ \wedge {\bf \bar{F}}_\times \wedge {\bf \bar{F}}_{\rm vx} \wedge {\bf \bar{F}}_\ell|} \cdot {\bf \bar{d}},  \\
{}^{(5)}d^3_{\rm GR, null} &=& \frac{{\bf \bar{F}}_+ \wedge {\bf \bar{F}}_\times \wedge {\bf \bar{F}}_{\rm vy} \wedge {\bf \bar{F}}_\ell}{|{\bf \bar{F}}_+ \wedge {\bf \bar{F}}_\times \wedge {\bf \bar{F}}_{\rm vy} \wedge {\bf \bar{F}}_\ell|} \cdot {\bf \bar{d}}. \\
\end{eqnarray}
If a theory that has scalar modes happens to be the right one, then there will be a signal in ${}^{(5)}d^1_{\rm GR, null}$ above. If there are vector modes,\index{vectorial modes} then they will show up in ${}^{(5)}d^2_{\rm GR,null}$ and/or ${}^{(5)}d^3_{\rm GR,null}$. 

Hayama and Nishizawa showed how to reconstruct the polarization modes in the case where the number of detectors is at least the number of modes, based on the null stream idea \cite{Hayama2012}. As an illustration, they reconstructed a simulated longitudinal mode\index{longitudinal mode} in Brans-Dicke theory.\index{Brans-Dicke theory} Such a mode might be emitted by a supernova explosion, in which radial velocities $v/c \sim 0.25$ are reached \cite{Fryer2006}. 

If a statistical excess is seen in one or more null streams, then one would like to match-filter them with template waveforms that allow for one or more alternative polarization states\index{polarization states} to obtain information about their physical content. Such waveform models were developed in the context of the (extended) parameterized post-Einsteinian (ppE) framework\index{parameterized post-Einsteinian framework} by Chatziioannou \emph{et al.} \cite{Chatziioannou2012}, and we refer the reader to that paper for details. The original ppE framework will be discussed below.

\section{Probing gravitational self-interaction}
\label{sec:self-interaction}

\subsection{The regime of late inspiral}
\label{subsec:inspiral}

Within GR, especially the inspiral\index{inspiral} part of the coalescence process has been modeled in great detail using the post-Newtonian (PN) formalism\index{post-Newtonian formalism} (see \cite{Blanchet2006} and references therein), in which quantities such as the conserved energy and flux are found as expansions in $v/c$, where $v(t)$ is a characteristic speed. During inspiral, the GW signals will carry a detailed imprint of the orbital motion. Indeed, the main contribution has a phase that is simply $2\Phi(t)$, with $\Phi(t)$ the orbital phase.\index{orbital phase} Thus, the angular motion of the binary is directly encoded in the waveform's phase, and assuming quasi-circular inspiral, the radial motion follows from the instantaneous angular frequency $\omega(t)=\dot{\Phi}(t)$ through the relativistic version of Kepler's Third Law. If there are deviations from GR, the different emission mechanism and/or differences in the orbital motion will be encoded in the phase of the signal waveform, allowing us to probe the strong-field dynamics of gravity.

In this section, we shall employ the usual post-Newtonian notation, in which ``$q$PN order", with $q$ an integer or half-integer, refers to contributions at $(v/c)^{2q}$ beyond leading order.  

Up to a reference phase, the orbital phase\index{orbital phase} takes the form \cite{Blanchet2002}
\begin{equation}
\Phi(v) = \left(\frac{v}{c}\right)^{-5} \sum_{n=0}^\infty \left[ \varphi_n + \varphi^{(l)}_n \ln\left(\frac{v}{c}\right) \right]\,\left(\frac{v}{c}\right)^n.
\label{phase}
\end{equation}
In general relativity, the coefficients $\varphi_n$ and $\varphi^{(l)}_n$ depend on the component masses $m_1$, $m_2$ and spins $\vec{S}_1$, $\vec{S}_2$ in a very specific way; these dependences are currently known up to $n=7$. The different PN terms in the phasing formula arise from non-linear multipole interactions as the wave propagates from the source's ``near zone", where gravitational fields are strong, to the ``far zone", where detection takes place. Specifically, the physical content of some of the contributions is as follows: 
\begin{itemize}
\item $\varphi_3$ and $\varphi_5$ encode the interaction of the total (Arnowitt-Deser-Misner, or ADM \cite{MTW}) mass-energy\index{ADM mass-energy} of the source with the quadrupole moment. The physical picture is that the quadrupolar waves scatter off the Schwarzschild curvature generated by the source. These contributions are referred to as gravitational wave ``tails"\index{gravitational wave tails}. One of the early proposals towards testing non-linear aspects of general relativity using gravitational waves was due to Blanchet and Sathyaprakash, who first discussed the possibility of measuring these tail effects \cite{Blanchet1994}. 
\item Spin-orbit interactions\index{spin-orbit interactions} also first make their appearance in $\varphi_3$, and the lowest-order spin-spin interactions\index{spin-spin interactions} occur in $\varphi_4$ \cite{Kidder1993}.
\item $\varphi_6$ includes the cubic non-linear interactions in the scattering of gravitational waves due to the ADM mass-energy\index{ADM mass-energy} of the system \cite{Blanchet1994}.
\end{itemize}
Thus, observations of these PN contributions would allow for penetrating tests of the non-linear structure of general relativity.

It is worth noting that with binary pulsars, one can only constrain the conservative sector of the orbital dynamics to 1PN order, and the dissipative sector to leading order; see, \emph{e.g.}, the discussion in \cite{Maggiore} and references therein. Hence, when it comes to $\Phi(t)$, these observations do not fully constrain the 1PN contribution. 
More generally, terms in (\ref{phase}) with $n > 0$ are only accessible with direct gravitational wave detection.

\subsection{The parameterized post-Einsteinian formalism}

By now there is a large body of literature on alternative theories to general relativity,\index{alternative theories of gravity} which will induce changes in the functional dependences of the $\varphi_n$, $\varphi_n^{(l)}$ on component masses and spins, or even introduce new powers of $v/c$ in the phase expression, Eq.~(\ref{phase}). For instance,
\begin{itemize}
\item The effect of a non-standard dispersion relation (\emph{e.g.} due to a non-zero graviton mass\index{graviton mass}) would accumulate over the large distances which the signal has to travel to reach the detector, and would be visible in $\varphi_2$. Solar system dynamics bound the graviton's Compton wavelength as $\lambda_g \gtrsim 10^{12}$ km. Second-generation detectors will improve on this by a factor of a few; Einstein Telescope\index{Einstein Telescope} will probe  $\lambda_g \gtrsim 10^{14}$ km, and LISA\index{LISA} $\lambda_g \gtrsim 10^{16}$ km \cite{Will1998,Will2004,Berti2005,Stavridis2009,Arun2009,Keppel2010,DelPozzo2011}.
\item Scalar-tensor theories\index{scalar-tensor theories} add a term $\varphi_{ST}\,(v/c)^{-7}$ to Eq.~(\ref{phase}), due to dipolar emission\index{dipolar emission}. In Brans-Dicke theory,\index{Brans-Dicke theory} one has a dimensionless parameter $\omega_{\rm BD}$ which leads to standard GR in the limit $\omega_{\rm BD} \rightarrow \infty$. The Solar system bound from the Cassini spacecraft is $\omega_{\rm BD} \gtrsim 40000$; LISA will improve on this by up to an order of magnitude \cite{Will2004,Berti2005,Yagi2010,Will1994,Scharre2002}.
\item A variable Newton constant adds a term $\varphi_{G(t)}\,(v/c)^{-13}$ \cite{Yunes2009b}, and extra dimensions\index{extra dimensions} can also have this effect \cite{Yagi2011}.
\item Quadratic curvature terms\index{quadratic curvature terms} in the Lagrangian modify $\varphi_4$ \cite{Stein2011}. The same is true of dynamical Chern-Simons theory\index{Chern-Simons theory} \cite{Yagi2012}. Here the second-generation detectors could place a bound on a dimensionful parameter of $\xi^{1/4} \lesssim \mathcal{O}(10-100)$ km, six to seven orders of magnitude better than the solar system constraint ($\xi^{1/4} \lesssim \mathcal{O}(10^8)$ km), and in this case also considerably better than LISA ($\xi^{1/4} \lesssim \mathcal{O}(10^5-10^6)$ km)!
\end{itemize}
Quadratic curvature terms arise in string theory compactifications \cite{Zweibach1985}, and dynamical Chern-Simons theory can be motivated both from string theory\index{string theory} \cite{Alexander2006} and loop quantum gravity\index{loop quantum gravity} \cite{Taveras2008}, and also arises in effective field theories of inflation\index{inflation} \cite{Weinberg2008}. Their effects on the phase at $(v/c)^4$ beyond leading order will only become visible when $v/c$ is large. This is the regime we will be interested in here.

Yunes and Pretorius established the so-called parameterized post-Einsteinian (ppE) framework\index{parameterized post-Einsteinian framework} as a way both to classify alternative theories of gravity,\index{alternative theories of gravity} and to provide template waveforms to search for violations of GR with gravitational wave detectors \cite{Yunes2009}. Their proposal involves both the phase and the amplitude of gravitational waves. However, since we are mostly concerned with second-generation detectors which for the expected stellar mass sources will not be very sensitive to changes in the amplitude \cite{VanDenBroeck2007}, we will focus on the phase. Instead of using the expression (\ref{phase}) for the inspiral\index{inspiral} phase, the authors of \cite{Yunes2009} proposed the following ansatz (again up to some reference phase):\footnote{The original proposal of \cite{Yunes2009} was formulated in the frequency domain and omitted logarithmic contributions, but the basic idea is the same.}
\begin{equation}
\Phi(v) = \sum_{n=0}^N \left[\phi_n + \phi_n^{(l)} \ln \left(\frac{v}{c}\right) \right] \left(\frac{v}{c}\right)^{b_n}.
\label{ppE_1}
\end{equation}
Here, the $b_n$ and $\phi_n$, $\phi^{(l)}_n$ are meant to be completely free parameters. The above phase  reduces to the one predicted by GR, Eq.~(\ref{phase}), for  $b_n = -5, -4, \ldots$ and when the phase coefficients have the standard dependences on component masses and spins: $\phi_n = \varphi_n(m_1, m_2, \vec{S}_1, \vec{S}_2)$, $\phi_n^{(l)} = \varphi_n^{(l)}(m_1, m_2, \vec{S}_1, \vec{S}_2)$. Yunes and Pretorius also showed how a variety of alternative theories of gravity\index{alternative theories of gravity} in the literature can be obtained by making appropriate choices for the $b_n$ and $\phi_n$. Now, in the case of second-generation detectors, the form (\ref{ppE_1}) may not be the most convenient one as far as data analysis is concerned. Indeed, even in the presence of a pure GR signal and using trial waveforms with the above phase, probability  distributions arising from measurements of $b_n$ and $\phi_n$ might peak at the correct values for very high SNRs, but probably not for signals at the threshold of detectability and in the presence of a considerable amount of noise, as is expected for most detections in second-generation observatories. 

Yunes and others calculated the phase for a great variety of alternative theories,\index{alternative theories of gravity} and in each case found the $b_n$ to be integer; see the examples and references above. It then makes sense to write
\begin{equation}
\Phi(v) = \sum_{n=-2}^N \left[ \phi_n + \phi_n^{(l)} \ln \left(\frac{v}{c}\right)  \right] \left(\frac{v}{c}\right)^{n-5},
\label{phaseintermediate}
\end{equation}
where we let the leading-order term be $(v/c)^{-7}$ to allow for dipolar emission.\index{dipolar emission} This time only the $\phi_n$ and $\phi_n^{(l)}$ are free parameters.

If there are too many free parameters to be determined, the measurement accuracy of \emph{all} of the parameters will be adversely affected, and we would still like to reduce the number of free $\phi_n$, $\phi_n^{(l)}$ in (\ref{phaseintermediate}). Alternative theories that have a non-zero $n = -2$ contribution to the phase, such as scalar-tensor theories,\index{scalar-tensor theories} can already be fairly well constrained using the electromagnetically observed binary pulsars  \cite{Freire2012}. With direct gravitational wave detection, the regime where we will be the most sensitive to GR violations is the one where $v/c$ is large, which is out of reach for other observational methods.  Hence we are mostly interested in new contributions to the phase with a power of $v/c$ greater than or equal to $-5$. For this reason, below we will set $\phi_{-2} = \phi_{-1} = 0$. 

\subsection{A generic test of general relativity with inspiraling compact binaries: the TIGER method\index{TIGER method}}
\label{subsec:TIGER}

\begin{figure}[htbp!]
\centering
\includegraphics[height=4cm]{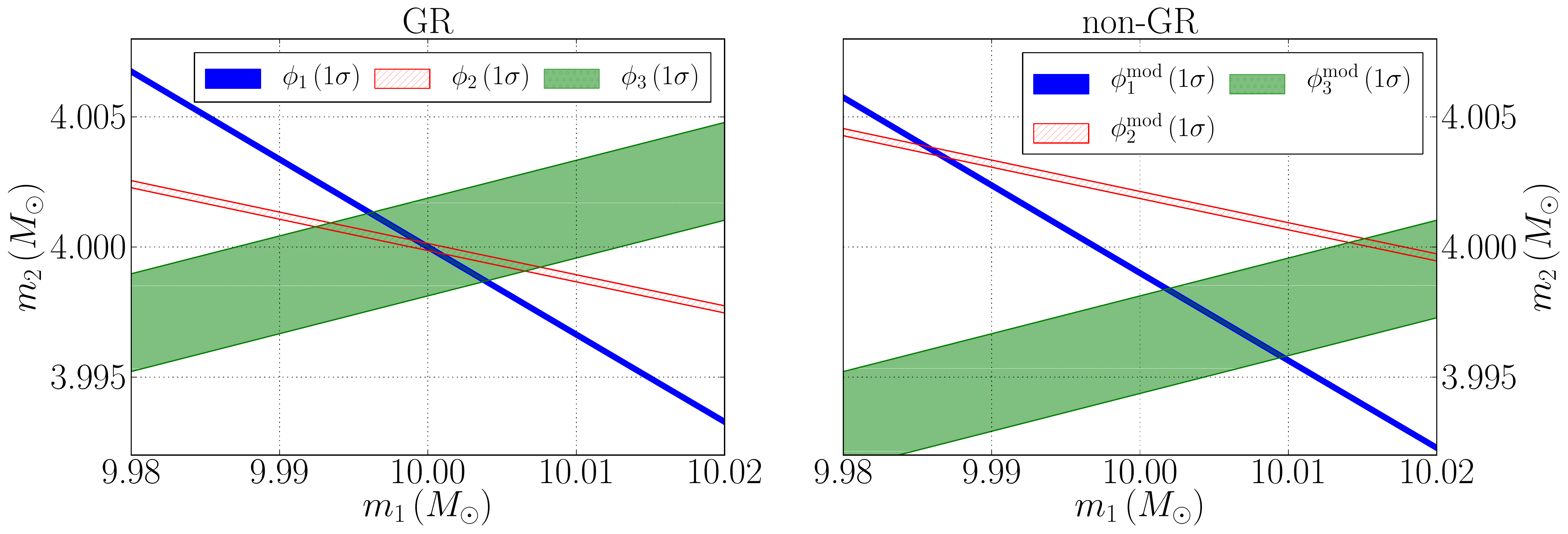}
\caption{A schematic illustration of how one might set up a very generic test of GR \cite{Arun2006a}. The plots show the regions in the plane of the component masses $(m_1,m_2)$ corresponding to the 1-$\sigma$ measurement uncertainties on the coefficients $(\phi_1,\phi_2,\phi_3)$. Left: If GR is correct, there will be a common region of overlap at the true values of the masses (here 10 and 4 $M_\odot$). Right: if there is a deviation from GR and one or more of the $\phi_n$ do not have the dependences on masses that GR predicts, then there will be a mismatch.}
\label{fig:params}
\end{figure}

In probing the strong-field dynamics, one would like to be sensitive to almost \emph{any} departure from general relativity, also through mechanisms that have yet to be envisaged. Hence what is needed is a test of GR that is as generic as possible. The possibility of such a test was first put forward by Arun \emph{et al.}~in \cite{Arun2006a}, and the idea is illustrated in Fig.~\ref{fig:params}. If for simplicity we assume that the component objects have zero spins, then the GR values of the coefficients $\phi_n$, $\phi^{(l)}_n$ in Eq.~(\ref{phaseintermediate}) only depend on the component masses $(m_1, m_2)$. Hence only two of them are independent, and tests of GR could be performed by comparing any three of them and checking for consistency.\footnote{Needless to say, this does not mean that a \emph{completely} generic test of GR is possible. In this picture, in principle there could be a GR violation which somehow still causes the error bands of any triplet of phasing coefficients to have a common region of overlap, but at the same, wrong component masses. See also the discussion in \cite{Vallisneri2012}.}  

In practice, it is more convenient to make use of \emph{Bayesian inference}\index{Bayesian inference}. This involves the comparison of two hypotheses, namely the GR hypothesis $\mathcal{H}_{\rm GR}$, and $\mathcal{H}_{\rm modGR}$ which posits that GR is violated. In the present context, $\mathcal{H}_{\rm GR}$ will be the hypothesis that  the $\phi_n$, $\phi_n^{(l)}$ depend on both masses and spins in the standard way. Ideally, $\mathcal{H}_{\rm modGR}$ would be the negation of $\mathcal{H}_{\rm modGR}$, but this is impossible in principle to evaluate, as one cannot check the observed phase against \emph{all} possible phase models that deviate from the GR family. Instead, we need to base our $\mathcal{H}_{\rm modGR}$ on a phase which allows for a finite-dimensional family of deviations. 

Inspired by \cite{Arun2006a}, we define $\mathcal{H}_{\rm GR}$ and $\mathcal{H}_{\rm modGR}$ as follows \cite{Li2012a,Li2012b}.
\begin{itemize}
\item $\mathcal{H}_{\rm GR}$ is the hypothesis that all the $\phi_n$, $\phi_n^{(l)}$ have the functional dependence on component masses and spins as predicted by GR.
\item $\mathcal{H}_{\rm modGR}$ is the hypothesis that \emph{one or more} of the  $\phi_n$, $\phi_n^{(l)}$ (without specifying which) do not have this functional dependence, but all others do.
\end{itemize}
Given a detected inspiral\index{inspiral} signal in a stretch of data $d$, the question is now how these hypotheses are to be evaluated.

Suppose we would like to compare two hypotheses $\mathcal{H}_A$ and $\mathcal{H}_B$. First, on each of them we can apply Bayes' theorem\index{Bayes' theorem} \cite{Jaynes}. For instance, for $\mathcal{H}_A$,
\begin{equation}
P(\mathcal{H}_A|d,I) = \frac{P(d|\mathcal{H}_A,I) P(\mathcal{H}_A|I)}{P(d|I)}.
\label{Bayes}
\end{equation}
Here $P(\mathcal{H}_A|d,I)$ is the \emph{posterior probability}\index{posterior probability} of the hypothesis $\mathcal{H}_A$ given the data $d$ and whatever additional information $I$ we may hold, $P(\mathcal{H}_A|I)$ is the \emph{prior probability}\index{prior probability} of the hypothesis, and $P(d|\mathcal{H}_A, I)$ is the \emph{evidence}\index{evidence} for $\mathcal{H}_A$, which can be written as
\begin{equation}
P(d|\mathcal{H}_A, I) = \int d\vec{\theta}\, p(d|\mathcal{H}_A, \vec{\theta}, I)\,p(\vec{\theta}|\mathcal{H}_A, I).
\label{evidenceintegral}
\end{equation}
In this expression, $p(\vec{\theta}|\mathcal{H}_A, I)$ is the prior probability density of the unknown parameter vector $\vec{\theta}$ within the model corresponding to $\mathcal{H}_A$, and $p(d|\mathcal{H}_A, \vec{\theta}, I)$ is the likelihood function\index{likelihood function} for the observation $d$, assuming the model $\mathcal{H}_A$ and given values of the parameters $\vec{\theta}$. 

The function $p(d|\mathcal{H}_A, \vec{\theta}, I)$ is what can be computed from the data. Let us assume that $\mathcal{H}_A$ corresponds to a particular gravitational wave signal model, $h_A(\vec{\theta}; t)$. In the output of a gravitational wave detector $d(t)$, the signal will be combined with detector noise $n(t)$:
\begin{equation}
d(t) = n(t) + h_A(\vec{\theta}; t). 
\label{detectoroutput}
\end{equation}
Let us assume that the noise is stationary and Gaussian; then its probability density distribution can be written as
\begin{equation}
p[n] = \mathcal{N}\,e^{-(n|n)/2}, 
\label{noisedistribution}
\end{equation}
where the square brackets in the left hand side indicate that $p[n]$ is a \emph{functional} of $n$, and $\mathcal{N}$ is a normalization factor. The inner product $(\,\cdot\, | \,\cdot\,)$ is defined as follows:
\begin{equation}
(a|b) = 4\,\Re \int_0^\infty \frac{\tilde{a}^\ast(f)\,b(f)}{S_n(f)},
\label{innerproduct}
\end{equation}
with $\tilde{a}(f)$, $\tilde{b}(f)$ the Fourier transforms of functions $a(t)$, $b(t)$. The quantity $S_n(f)$ is called the \emph{noise power spectral density}; comparing with Eq.~(\ref{noisedistribution}), we see that it is essentially the variance of the noise as a function of frequency. Eqns.~(\ref{detectoroutput}), (\ref{noisedistribution}), and (\ref{innerproduct}) motivate the following form for the likelihood\index{likelihood function}
$p(d|\mathcal{H}_A, \vec{\theta},  I)$:
\begin{equation}
p(d|\mathcal{H}_A, \vec{\theta},  I) = \mathcal{N}\,e^{-(d - h_A(\vec{\theta}) | d - h_A(\vec{\theta}) )/2}.
\label{likelihoodfunction}
\end{equation} 
Indeed, when subtracting the signal from the detector output, the expectation is that only stationary, Gaussian noise remains.

Using Eq.~(\ref{Bayes}) for both $\mathcal{H}_A$ and $\mathcal{H}_B$, one can construct an \emph{odds ratio}\index{odds ratio}
\begin{equation}
O^A_B \equiv \frac{P(\mathcal{H}_A|d, I)}{P(\mathcal{H}_B|d, I)} = \frac{P(\mathcal{H}_A|I)}{P(\mathcal{H}_B|I)}
\frac{P(d|\mathcal{H}_A, I)}{P(d|\mathcal{H}_B, I)},
\end{equation}
where $P(\mathcal{H}_A|I)/P(\mathcal{H}_B|I)$ is the \emph{prior odds}\index{prior odds} of the two hypotheses, \emph{i.e.}~the relative confidence we assign to the models before any observation has taken place. The ratio of evidences\index{evidence} is called the \emph{Bayes factor},\index{Bayes factor} which can be computed from the data by using Eqns.~(\ref{evidenceintegral}) and (\ref{likelihoodfunction}) for hypotheses $\mathcal{H}_A$ and $\mathcal{H}_B$:
\begin{equation}
B^A_B \equiv \frac{P(d|\mathcal{H}_A, I)}{P(d|\mathcal{H}_B, I)}.
\label{BAB}
\end{equation}

In the present context, the odds ratio\index{odds ratio} of interest is
\begin{equation}
O^{\rm modGR}_{\rm GR} = \frac{P(\mathcal{H}_{\rm modGR}|d, I)}{P(\mathcal{H}_{\rm GR}|d, I)} = \frac{P(\mathcal{H}_{\rm modGR}|I)}{P(\mathcal{H}_{\rm GR}|I)}
\frac{P(d|\mathcal{H}_{\rm modGR}, I)}{P(d|\mathcal{H}_{\rm GR}, I)}.
\label{OmodGRGR}
\end{equation}
The evidence\index{evidence} $P(d|\mathcal{H}_{\rm GR}, I)$ is computed by considering a large number of GR waveforms with different parameters $\vec{\theta}$ to map out the likelihood function\index{likelihood function} $p(d|\mathcal{H}_{\rm GR}, \vec{\theta}, I)$, Eq.~(\ref{likelihoodfunction}), which is then substituted into Eq.~(\ref{evidenceintegral}). However, the way $\mathcal{H}_{\rm modGR}$ is formulated, there is no waveform family associated with it, as there is no waveform model in which ``one or more" of the $\phi_n$, $\phi^{(l)}_n$ are different from their GR predictions. 

To address this issue, we introduce the following \emph{auxiliary hypotheses}:\footnote{With minor abuse of notation, from now on we let $\{\phi_0, \phi_2, \ldots,  \phi_M\}$ be the set of $M=10$ coefficients that are currently known from post-Newtonian calculations, including the ``logarithmic" ones $\phi^{(l)}_n$.}
\begin{quote}
$H_{i_1 i_2 \ldots i_k}$ is the hypothesis that the phasing coefficients $\phi_{i_1}, \phi_{i_2}, \dots, \phi_{i_k}$ do \emph{not} have the functional dependence on masses and spins as predicted by GR, but all other coefficients $\phi_j$, $j \notin \{i_1, i_2, \ldots, i_k\}$ \emph{do} have the dependence as in GR.
\end{quote}
Thus, for example, $H_{12}$ is the hypothesis that $\phi_1$ and $\phi_2$ deviate from their GR values, but all other coefficients are as in GR. With each of the hypotheses above, we can associate a waveform model that can be used to test it. Let $\vec{\theta} = \{m_1, m_2, \vec{S}_1, \vec{S}_2, \ldots \}$ be the parameters occurring in the GR waveform, where $m_1$, $m_2$ are the component masses and $\vec{S}_1$, $\vec{S}_2$ the component spins; other parameters include the orientation of the orbital plane with respect to the line of sight, sky position, and distance. Then $H_{i_1 i_2 \ldots i_k}$ is tested by a waveform in which the independent parameters are
\begin{equation}
\{\vec{\theta}, \phi_{i_1}, \phi_{i_2}, \ldots, \phi_{i_k} \},
\end{equation}
\emph{i.e.} the coefficients $\{\phi_{i_1}, \phi_{i_2}, \ldots, \phi_{i_k}\}$ are allowed to vary freely in addition to the other parameters. 

The hypothesis we are really interested in is $\mathcal{H}_{\rm modGR}$ above, which posits that one or more of the $\phi_i$ differ from their GR values, without specifying which. But this corresponds to the logical ``or" of the auxiliary hypotheses:
\begin{equation}
\mathcal{H}_{\rm modGR} = \bigvee_{i_1 < i_2 < \ldots < i_k; k \leq N_T} H_{i_1 i_2 \ldots i_k}.
\label{HmodGR}
\end{equation}
Note that in practice, it will not be possible for computational reasons to consider all possible subsets of even the 10 known phasing coefficients; hence we limit ourselves to the subsets of $\{\phi_1, \phi_2, \ldots, \phi_{N_T}\}$, where $N_T \leq 10$ is mainly set by computational resources. We will call the latter our \emph{testing coefficients}.

To illustrate how the auxiliary hypotheses allow us to compute the odds ratio\index{odds ratio} $\mathcal{O}^{\rm modGR}_{\rm GR}$ of Eq.~(\ref{OmodGRGR}), let us consider the case of just two testing coefficients, $\{\phi_1, \phi_2\}$. Then 
\begin{equation}
\mathcal{H}_{\rm modGR} = H_1 \vee H_2 \vee H_{12},
\end{equation}
and the odds ratio becomes
\begin{equation}
O^{\rm modGR}_{\rm GR} = \frac{P(H_1 \vee H_2 \vee H_{12}|d, I)}{P(\mathcal{H}_{\rm GR}|d, I)}.
\end{equation}
Now, the hypotheses $H_1$, $H_2$, and $H_{12}$ are \emph{logically disjoint}: the ``and" of any two of them is false. Indeed, in $H_1$, $\phi_2$ takes the GR value, but in $H_2$ it differs from it, as it does in $H_{12}$. Similarly, in $H_2$, $\phi_1$ takes the GR value, but it differs from it in $H_1$ and in $H_{12}$. This implies
\begin{equation}
P(H_1 \vee H_2 \vee H_{12}|d, I) = P(H_1|d,I) + P(H_2|d,I) + P(H_{12}|d,I)
\end{equation}
and hence
\begin{equation}
O^{\rm modGR}_{\rm GR} = \frac{P(H_1|d, I)}{P(\mathcal{H}_{\rm GR}|d, I)} + \frac{P(H_2|d, I)}{P(\mathcal{H}_{\rm GR}|d, I)} + \frac{P(H_{12}|d, I)}{P(\mathcal{H}_{\rm GR}|d, I)}.
\end{equation}
Using Bayes' theorem\index{Bayes' theorem} (\ref{Bayes}) on each term, we get
\begin{equation}
O^{\rm modGR}_{\rm GR} = \frac{P(H_1|I)}{P(\mathcal{H}_{\rm GR}|I)} B^1_{\rm GR} + \frac{P(H_2|I)}{P(\mathcal{H}_{\rm GR}|I)} B^2_{\rm GR} + \frac{P(H_{12}|I)}{P(\mathcal{H}_{\rm GR}|I)} B^{12}_{\rm GR},
\label{Oddsintermediate}
\end{equation}
where the Bayes factors $B^1_{\rm GR}$, $B^2_{\rm GR}$, and $B^{12}_{\rm GR}$ are given by
\begin{equation}
B^1_{\rm GR} = \frac{P(d |H_1, I)}{P(d|\mathcal{H}_{\rm GR}, I)},\,\,\,\,\,\,\,
B^2_{\rm GR} = \frac{P(d |H_2, I)}{P(d|\mathcal{H}_{\rm GR}, I)},\,\,\,\,\,\,\,
B^{12}_{\rm GR} = \frac{P(d |H_{12}, I)}{P(d|\mathcal{H}_{\rm GR}, I)}.
\label{Bayesfactors}
\end{equation}
These can be computed from the data, as explained in the discussion leading up to Eq.~(\ref{BAB}). However, a choice will have to be made for the relative prior odds $P(H_1|I)/P(\mathcal{H}_{\rm GR}|I)$, $P(H_2|I)/P(\mathcal{H}_{\rm GR}|I)$, and $P(H_{12}|I)/P(\mathcal{H}_{\rm GR}|I)$. If one believed that the graviton has mass, then a deviation in $\phi_2$ would be the thing to look for, and the auxiliary hypothesis $H_2$ should have more weight than either $H_1$ or $H_{12}$. On the other hand, one's favorite alternative theory might predict a violation in $\phi_1$ instead, in which case $H_1$ should have more prior weight. Or, one might expect a GR violation to affect \emph{all} phasing coefficients at the same time, so that $H_{12}$ is \emph{a priori} preferred. The method presented here is meant to find \emph{generic} deviations from GR, with no preference for any particular alternative theory; consequently, we set
\begin{equation}
\frac{P(H_1|I)}{P(\mathcal{H}_{\rm GR}|I)} = \frac{P(H_2|I)}{P(\mathcal{H}_{\rm GR}|I)} = \frac{P(H_{12}|I)}{P(\mathcal{H}_{\rm GR}|I)}.
\label{relativepriorodds}
\end{equation}
We will also need to specify the \emph{overall} prior odds for $\mathcal{H}_{\rm modGR}$ against $\mathcal{H}_{\rm GR}$. Here we simply set
\begin{equation}
\frac{P(\mathcal{H}_{\rm modGR}|I)}{P(\mathcal{H}_{\rm GR}|I)} = \frac{P(H_1 \vee H_2 \vee H_{12}|I)}{P(\mathcal{H}_{\rm GR}|I)} = \alpha,
\label{overallpriorodds}
\end{equation}
where the constant $\alpha$ will end up being an unimportant overall scaling factor in the odds ratio.\index{odds ratio} Eqns.~(\ref{relativepriorodds}), (\ref{overallpriorodds}) imply 
\begin{equation}
\frac{P(H_1|I)}{P(\mathcal{H}_{\rm GR}|I)} = \frac{P(H_2|I)}{P(\mathcal{H}_{\rm GR}|I)} = \frac{P(H_{12}|I)}{P(\mathcal{H}_{\rm GR}|I)} = \frac{\alpha}{3},
\end{equation}
and, except for the overall factor $\alpha$, the final expression for the odds ratio reduces to a straightforward average of the Bayes factors for the auxiliary hypotheses against GR:
\begin{equation}
O^{\rm modGR}_{\rm GR} = \frac{P(\mathcal{H}_{\rm modGR}|d,I)}{P(\mathcal{H}_{\rm GR}|d,I)}  = \frac{\alpha}{3}\,\left[B^1_{\rm GR} + B^2_{\rm GR} + B^{12}_{\rm GR}\right].
\end{equation}
Thus, although there is no waveform model with which to directly test the hypothesis $\mathcal{H}_{\rm modGR}$, thanks to the auxiliary hypotheses it is nevertheless possible to compute its posterior probability\index{posterior probability} relative to that of GR.

In the above example we used only two testing parameters, but in practice one will want to have more. With $N_T$ testing parameters $\{\phi_1, \ldots, \phi_{N_T}\}$ and making similar choices to Eqns.~(\ref{relativepriorodds}), (\ref{overallpriorodds}), the odds ratio\index{odds ratio} will again be proportional to an average of the Bayes factors for the auxiliary hypotheses against GR \cite{Li2012a}:
\begin{equation}
O^{\rm modGR}_{\rm GR} = \frac{\alpha}{2^{N_T} - 1}\,\sum_{k=1}^{N_T} \sum_{i_1 < i_2 < \ldots < i_k} B^{i_1 i_2 \ldots i_k}_{\rm GR}.
\label{OddsSingleSource}
\end{equation}

Combining data from multiple observed inspiral\index{inspiral} events will make for a far more robust test of GR compared to using just one detection. Suppose one has $\mathcal{N}$ independent detections in stretches of data $d_1, d_2, \ldots, d_{\mathcal{N}}$. Assuming these to be independent, it is not difficult to show that the odds ratio\index{odds ratio} for the ``catalog" of detections as a whole takes the form \cite{Li2012a}
\begin{eqnarray}
\mathcal{O}^{\rm modGR}_{\rm GR} &=& \frac{P(\mathcal{H}_{\rm modGR}|d_1, d_2, \ldots, d_\mathcal{N}, I)}{P(\mathcal{H}_{\rm GR}|d_1, d_2, \ldots, d_\mathcal{N}, I)} \nonumber\\ 
&=& \frac{\alpha}{2^{N_T}-1} \sum_{k=1}^{N_T} \sum_{i_1 < i_2 < \ldots <i_k} \prod_{A = 1}^\mathcal{N} {}^{(A)}B^{i_1 i_2 \ldots i_k}_{\rm GR},
\label{OddsCatalog}
\end{eqnarray}
\emph{i.e.}, for each auxiliary hypothesis, one multiplies together all the Bayes factors against GR for individual sources, ${}^{(A)}B^{i_1 i_2 \ldots i_k}_{\rm GR}$, after which one takes the average over all these hypotheses.

The algorithm described here was developed by Li \emph{et al.} \cite{Li2012a,Li2012b}. It has been dubbed the TIGER  method\index{TIGER method} (``Test Infrastructure for GEneral Relativity"), and a hands-on data analysis pipeline for use on the upcoming detections in Advanced LIGO\index{Advanced LIGO} and Virgo\index{Advanced Virgo} data has been developed based on this idea. It has a number of benefits:
\begin{itemize}
\item Unlike previous Bayesian treatments such as \cite{Cornish2011,Gossan2012}, it addresses the question ``Do \emph{one or more} testing parameters deviate from their GR values?", as opposed to ``Do \emph{all} of them deviate?". Bayesian analysis naturally includes the idea of Occam's Razor\index{Occam's Razor} in a quantitative way, and if the full non-GR model happens to have too many free parameters then one will be penalized for it \cite{Jaynes}.
\item It is well-suited to a situation where most sources are near the threshold of detectability. As shown in \cite{Li2012a}, if a GR violation is small, the Bayes factor for the ``correct" auxiliary hypothesis (if any) will not always make the largest contribution to the odds ratio,\index{odds ratio} as detector noise can obfuscate the precise nature of the GR violation. Even then, the GR hypothesis will typically be disfavored compared to one or more of the other auxiliary hypotheses, causing the GR violation to be detected after all.
\item In combining information from multiple sources, it is not necessary that a GR violation manifests itself in the same way from one source to another. A deviation from GR could depend on mass, and on whatever additional charges might be present in the correct theory of gravity. However, in the above, the same ``yes/no" question is asked for every source, and evidence\index{evidence} for or against GR is allowed to build up as more and more sources get added.
\item The method is not restricted to just the inspiral\index{inspiral} phase. It could equally well be applied to ringdown (as discussed below), or for that matter to alternative polarization states.\index{polarization states} All that is needed is a convenient parameterization of possible deviations from GR, such as provided by (generalizations of) the parameterized post-Einsteinian formalism.
\end{itemize}

\subsection{Accuracy in probing the strong-field dynamics with second-generation detectors}

Let us consider some examples to gauge how sensitive the TIGER method\index{TIGER method} will be for particular (though heuristic) violations of GR, with the network of Advanced LIGO\index{Advanced LIGO} and Virgo detectors.\index{Advanced Virgo} In order to do this, one can produce simulated stationary, Gaussian detector noise, whose power spectral density (essentially the variance of the noise as a function of frequency) is in accordance with predictions for the Advanced LIGO and Virgo interferometers in their final configurations, projected for the 2019-2021 time frame \cite{LIGO,Virgo}.  Simulated signals can be added to this simulated noise. 

First we consider binary neutron stars.\index{binary neutron stars} For such sources, the inspiral\index{inspiral} signal ends at high frequencies, and to good approximation one can assume that only this part of the coalescence process is visible in the frequency band where the detectors are sensitive. Moreover, in BNS systems the dimensionless intrinsic spins of the components are expected to be small: $c J/(G m^2) \lesssim 0.05$ \cite{OShaughnessy2008}, with $J$ the spin and $m$ the component mass. Finally, for most of the inspiral, neutron stars can be treated as point particles; finite size and matter effects will only be important at relatively high frequencies where the detectors are not very sensitive \cite{Hinderer2010}. 
Thus, binary neutron star inspirals are relatively clean systems whose GW emission can be described by fairly simple waveform models \cite{BIOPS}. Indeed, a hands-on data analysis pipeline which starts from raw detector data and computes $\mathcal{O}^{\rm modGR}_{\rm GR}$ has already been developed.


To have a fair assessment of how the method will perform with second-generation detectors, the simulated BNS sources will have to be distributed in an astrophysically realistic way. We will assume the component masses to be uniform in the interval $[1, 2]\,M_\odot$. The normal to the inspiral plane, and the sky position, are taken from a uniform distribution on the sphere, and sources are distributed uniformly in volume. Distances are between 100 and 400 Mpc; the former is the distance within which one can realistically expect one inspiral event every two years, and the latter is approximately the largest distance at which an optimally oriented and positioned system is still visible with Advanced LIGO\index{Advanced LIGO} \cite{ratespaper}. Finally, many simulated catalogs of 15 sources each are produced.

We will be interested in GR violations that affect contributions to the phase (\ref{phaseintermediate}) with $n > 0$; as mentioned before, we expect novel effects to show up for large $v/c$. Therefore, let us choose as testing coefficients the set $\{\phi_1, \phi_2, \phi_3\}$, leading to $2^3 - 1 = 7$ auxiliary hypotheses that need to be compared with $\mathcal{H}_{\rm GR}$ in order to compute the odds ratio\index{odds ratio} $\mathcal{O}^{\rm modGR}_{\rm GR}$. 

\emph{A priori}, one would expect $\mathcal{O}^{\rm modGR}_{\rm GR} > 1$ if GR is violated, and $\mathcal{O}^{\rm modGR}_{\rm GR} < 1$ if it is correct. However, detector noise can mimick GR violations, so that occasionally one will have $\mathcal{O}^{\rm modGR}_{\rm GR} > 1$ even when GR is in fact correct. To deal with this, one can compute odds ratios\index{odds ratio} for a large number of catalogs of simulated sources whose emission is in accordance with GR, but having different parameter values within the above ranges, and see how the odds ratio ends up being distributed. For a given kind of GR violation, one can similarly construct an odds ratio distribution for catalogs of  simulated sources. If the non-GR distribution has significant overlap with the GR distribution, then the particular GR violation considered would not necessarily be detected with great confidence. If, on the other hand, the two distributions are perfectly separated then the GR violation will be incontrovertibly detectable. 

\begin{figure}[htbp!]
\centering
\includegraphics[height=5.5cm]{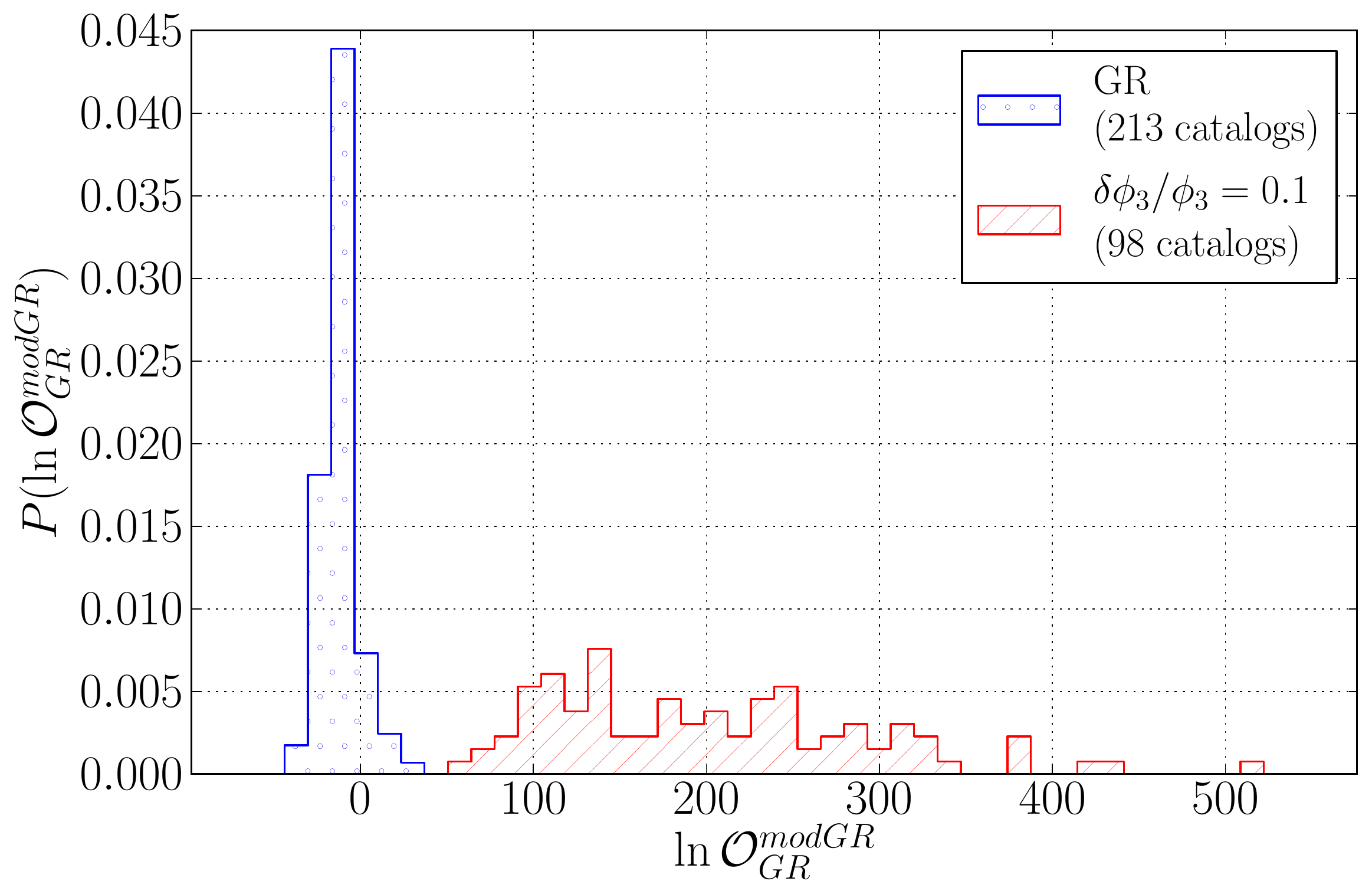}
\includegraphics[height=5.5cm]{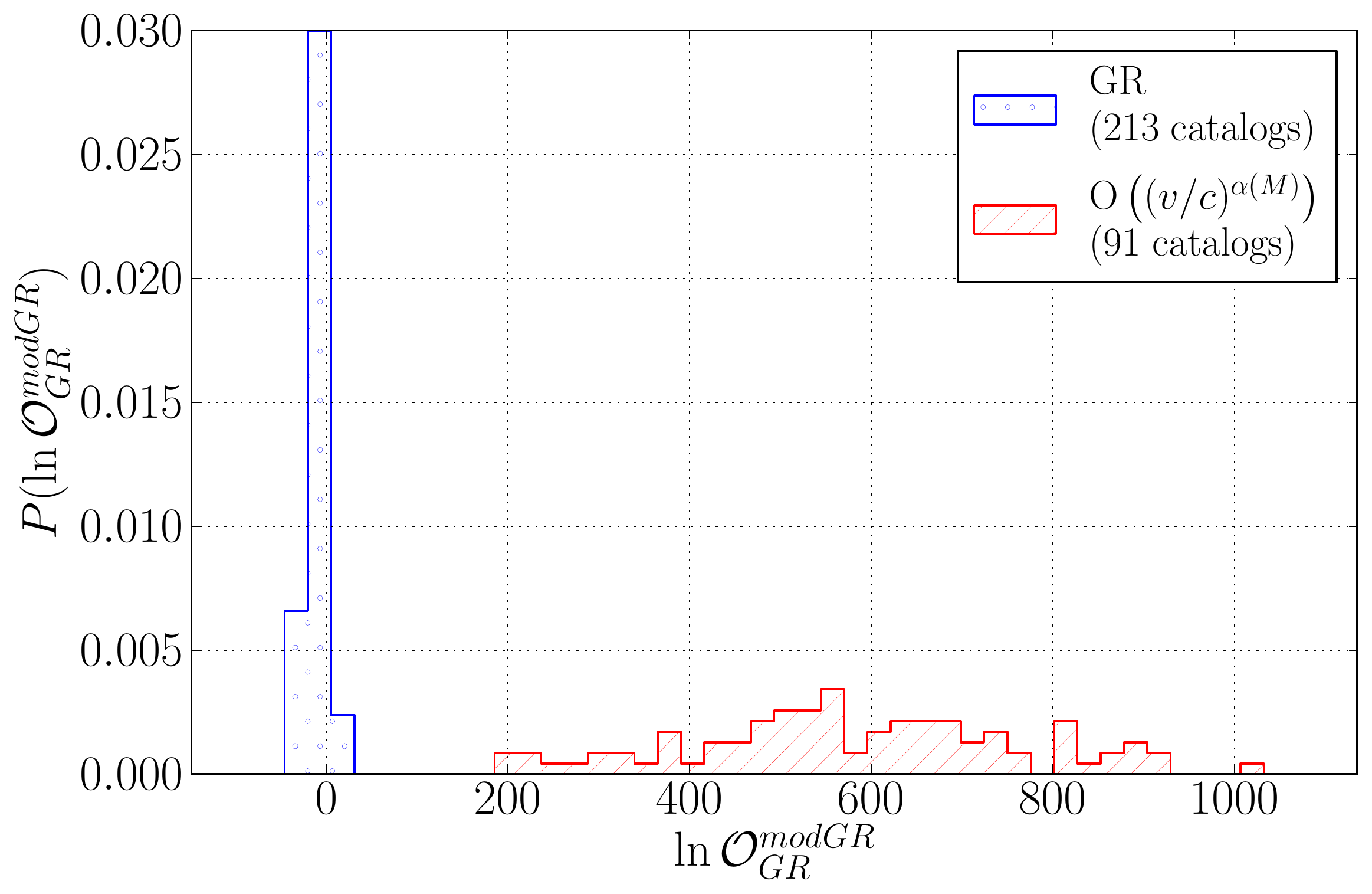}
\caption{Top: Distributions of log odds ratios for many catalogs of 15 BNS sources each, with and without a 10\% shift in $\phi_3$, the 1.5PN phasing coefficient which contains the leading-order non-linearities of GR related to tail effects \cite{Li2012a}. The blue, dotted distribution is for sources with pure GR emission; we see that mostly $\ln \mathcal{O}^{\rm modGR}_{\rm GR} < 0$, although noise will occasionally mimick a GR violation, causing the tail towards positive $\ln \mathcal{O}^{\rm modGR}_{\rm GR}$. The red, dashed distribution is for sources with the 10\% shift at 1.5PN. The two distributions are perfectly separated, indicating that a violation of this type and magnitude will easily be detected. Bottom: The same for a violation which does not manifest itself as a simple shift in one or more of the phasing coefficients (see the main text for details), but $ \mathcal{O}^{\rm modGR}_{\rm GR}$ is still computed in exactly the same way as before \cite{Li2012b}. Here too, there is complete separation between GR and non-GR.}
\label{fig:deviations}
\end{figure}

As we have seen, non-linearities related to ``tail" effects first show up at 1.5PN order, \emph{i.e.} in $\phi_3$, so that this contribution is of particular interest. In order to gauge how sensitive the method might be to GR violations at this order, in \cite{Li2012a} a constant relative shift in $\phi_3$ was considered: $\phi_3 = (1 + \delta\chi_3)\,\phi^{\rm GR}_3$. This was compared with the GR case, and it was found that for $\delta\chi_3 = 0.1$, there is complete separation between the odds ratio\index{odds ratio} distributions for the GR and non-GR catalogs, so that a violation of this kind and size would certainly be discovered. This is shown in the top panel of Fig.~\ref{fig:deviations}.

What if a deviation from GR does not manifest itself as simple shifts in the phase coefficients? Naively one might think that a more general phase model as in Eq.~(\ref{ppE_1}) would then be needed to uncover the GR violation. To show that this is not the case, Li \emph{et al.} \cite{Li2012b} considered a heuristic violation of the form 
\begin{equation}
\Phi^{\rm GR}(v) \rightarrow \Phi^{\rm GR}(v) + \beta\,\left(\frac{v}{c}\right)^{-6 + M/M_\odot}, 
\label{veryanomalous}
\end{equation}
with $\Phi^{\rm GR}(v)$ the GR phase. The prefactor $\beta$ was chosen to be of the same order as the $\phi_n$ predicted by GR (see \cite{Li2012b} for details), and $M$ is the total mass, so that the power of $v/c$ in the extra term varies from effectively 0.5PN to effectively 1.5PN within the BNS mass range considered. However, in order to compute $\mathcal{O}^{\rm modGR}_{\rm GR}$, the phase model used was still that of Eq.~(\ref{phaseintermediate}) with integer $n$, and the testing parameters were $\{\phi_1, \phi_2, \phi_3\}$, as before. As shown in the bottom panel of Fig.~\ref{fig:deviations}, also in this eventuality the GR hypothesis $\mathcal{H}_{\rm GR}$ will be disfavored compared with one or more of the $H_{i_1 i_2 \ldots  i_k}$. The separation between GR and non-GR source catalogs is complete. The odds ratio\index{odds ratio} $\mathcal{O}^{\rm modGR}_{\rm GR}$ indeed provides a Bayesian realization of the basic idea sketched in Fig.~\ref{fig:params}, inspired by Arun \emph{et al.} \cite{Arun2006a}.

\subsection{Binary neutron stars versus binary black holes}

As mentioned before, in the case of binary neutron stars,\index{binary neutron stars} it is mostly only the inspiral\index{inspiral} part that is within the sensitivity band of the detector, so that we do not have to worry about the messy merger process. Finite size and matter effects mostly appear at high frequencies, where they have little impact. Neutron stars in binaries are expected to be relatively slowly spinning, and also this aspect can be dealt with. Consequently, a simple waveform model can be used for which data analysis algorithms are sufficiently fast, and a full data analysis pipeline for testing GR with BNS signals is already in place.

\begin{figure}[htbp!]
\centering
\includegraphics[height=6cm]{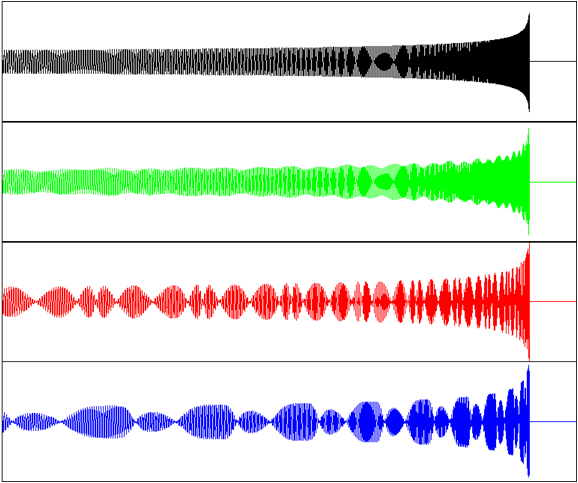}
\caption{For systems with no spins, the frequency and amplitude of gravitational waveforms increases in a steady ``chirp" (top). If there are significant spins which are not aligned with each other and the orbital angular momentum, then one has precession of the orbital plane, causing modulation of the amplitude and frequency. (Courtesy B.S.~Sathyaprakash.)}
\label{fig:modulation}
\end{figure}

The situation is quite different for binary black holes.\index{binary black holes} The frequency at which the inspiral terminates is roughly $c^3/(6^{3/2} \pi G M)$, with $M$ the total mass. For a BBH with component masses of $(10,10)\,M_\odot$, this is approximately 220 Hz, close to the frequency of $\sim 150$ Hz where the detectors will be the most sensitive. Thus, the merger part of the signal, which is still not well modeled, will play a major role. Moreover, astrophysical black holes are expected to be fast-spinning, with dimensionless spins $c J/(G m^2) = 0.3 - 0.99$ \cite{OShaughnessy2005}. If the spins are not aligned with each other and the orbital angular momentum, then all three of these vectors will undergo precession during the inspiral phase \cite{Apostolatos1994,Kidder1995}. Since the unit vector $\hat{L}$ in the direction of orbital angular momentum is also the unit normal to the inspiral plane, the latter will undergo precession, and in extreme cases even a tumbling motion. This behavior is imprinted onto the gravitational wave emission through modulation of both the amplitude and the frequency of the waveform, as shown in Fig.~\ref{fig:modulation}. The rich dynamics that is unleashed in this way makes binary black holes far more interesting systems to study, but the more complicated signals also make the data analysis problem a great deal more difficult.
 
Large-scale numerical simulations provide us with accurate waveform models \cite{Grandclement2009}, but they take a long time to compute and can not be used in data analysis, where many thousands of trial waveforms need to be compared with the data to arrive at accurate parameter estimation. On the other hand, semi-analytic inspiral-merger-ringdown\index{inspiral} waveforms are under construction, which roughly fall into two categories. In the \emph{Effective One-Body} formalism,\index{Effective One-Body formalism} the inspiral includes part of the final plunge, and a ringdown waveform can be ``stitched" to it; the waveform as a whole can then be further ``tuned" against numerical results  \cite{Damour2009}. There is also a variety of phenomenological waveform models which are similarly improved using numerical predictions \cite{Ajith2007,Ajith2011,Sturani2010}. These are achieving matches $\gtrsim 0.99$ with numerically predicted signals; however, so far the only inspiral-merger-ringdown waveform with fully precessing spins is the one of Refs.~\cite{Sturani2010}, which has been tuned against only a limited number of numerical waveforms. 

\begin{figure}[htbp!]
\centering
\includegraphics[height=5.5cm]{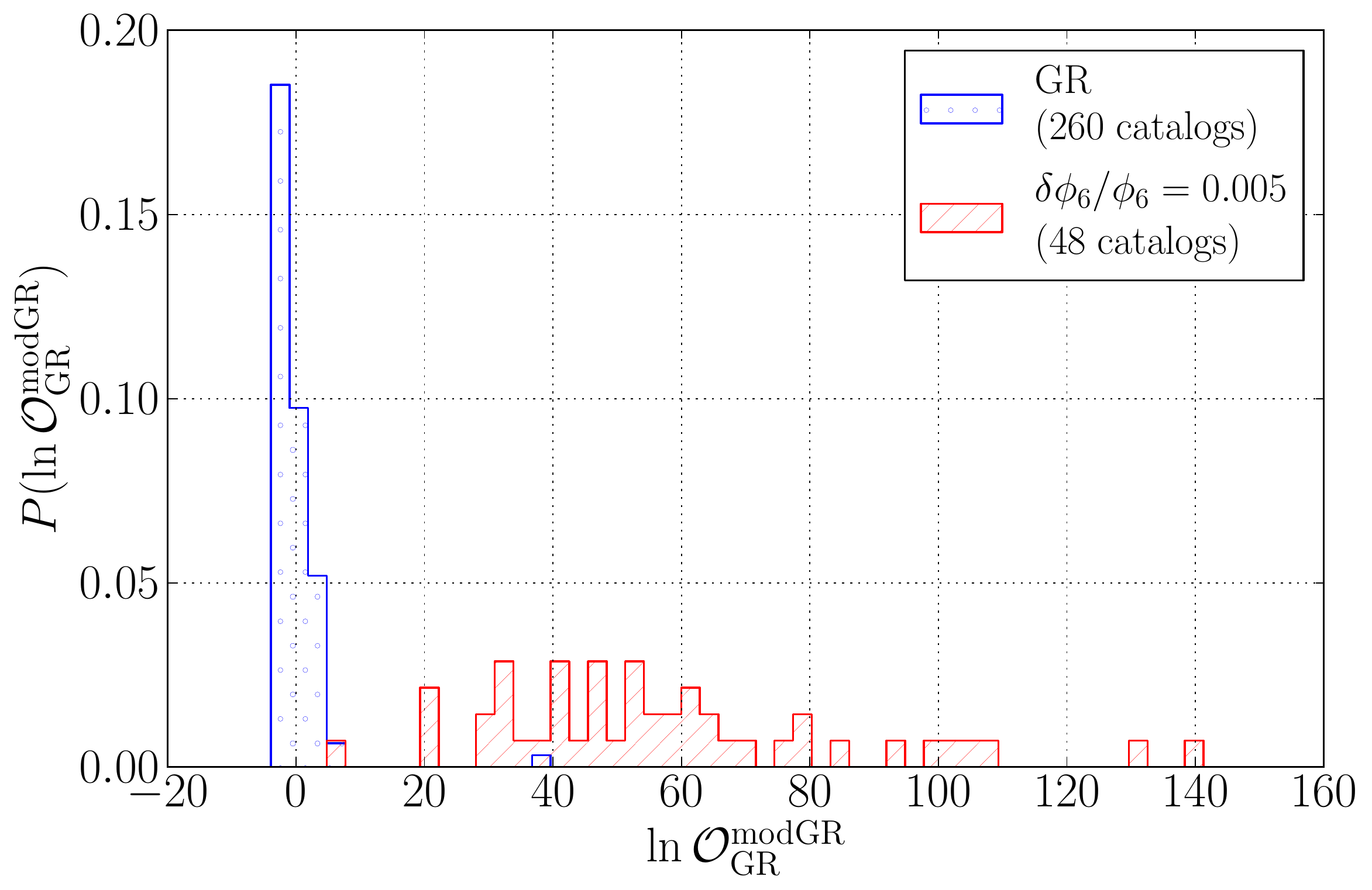}
\caption{An estimate by the authors of \cite{Li2012a,Li2012b} of how well one would be able to probe deviations from GR at high post-Newtonian order with binary black holes. As before, the blue dotted histogram is for catalogs of GR sources. The red dashed histogram is for sources with a 0.5\% deviation at 3PN order, where cubic non-linear self-interaction of the gravitational field appears. The testing coefficients were $\{\phi_1,\phi_2,\phi_3,\phi_4\}$ and hence did not include the parameter $\phi_6$ where the GR violation actually occurs; nevertheless, there is near-complete separation between GR and non-GR.}
\label{fig:IMRPhenom}
\end{figure}

Because of these difficulties, TIGER\index{TIGER method} can not yet be extended for use on BBH signals. However, the authors of \cite{Li2012a,Li2012b} made some rough estimates of what might be achievable once appropriate waveform models are available. Fig.~\ref{fig:IMRPhenom} shows the log odds ratio distributions for catalogs of simulated BBH signals with GR emission, and with a 0.5\% shift in $\phi_6$, which encodes the cubic non-linear interactions of the scattering of gravitational wave tails\index{gravitational wave tails} by the ADM mass-energy.\index{ADM mass-energy} In both cases, the inspiral-merger-ringdown\index{inspiral} waveform model used was the aligned-spin approximant of \cite{Ajith2011}, and the set of testing parameters consisted of (the analogs of) only $\{\phi_1,\phi_2,\phi_3,\phi_4\}$.\footnote{Note that the number of auxiliary hypotheses $H_{i_1 \ldots i_k}$ grows with the number of testing parameters $N_T$ as $2^{N_T}-1$, and the data analysis problem can become computationally very costly if too many are used.} Hence the the coefficient containing the deviation was not among the testing parameters. However, even though both $\mathcal{H}_{\rm GR}$ and the $H_{i_1 \ldots i_k}$ are inconsistent with the signal, the waveform models of the latter have more freedom and can arrive at a closer fit, causing the GR hypothesis to be disfavored. And indeed, there is near-complete separation between GR and non-GR! On the other hand, precessing spins are bound to affect these results, in a way that is as yet unknown.

\section{Testing the no hair theorem}
\label{sec:nohair}

In Newtonian theory, the gravitational potential $\Phi$ caused by a body with density $\rho$ satisfies
\begin{eqnarray}
\nabla^2 \Phi &=& 4\pi G \rho\,\,\,\,\,\,\,\mbox{in the interior}, \\
\nabla^2 \Phi &=& 0\,\,\,\,\,\,\,\,\,\,\,\,\,\,\,\,\,\,\,\mbox{in the exterior}.
\end{eqnarray}
In the exterior, $\Phi$ can be expanded as
\begin{equation}
\Phi = - G \sum_{l, m} \frac{M_{lm}}{r^{l+1}} Y_{lm}(\theta,\phi),
\end{equation}
and the \emph{multipole moments}\index{multipole moments} $M_{lm}$ are obtained by demanding consistency between the interior and exterior solutions. For axially symmetric objects, only terms with $m=0$ contribute. The lowest-order multipole, $M_{00}$, is just the total mass of the body. By appropriately choosing the center of the coordinate system used, one can set $M_{10} = 0$. The next non-trivial multipole moment is $M_{20}$, the quadrupole moment, which has dimensions $M L^2$. The set of all multipole moments uniquely determines the shape of the potential $\Phi$, and by measuring them one can study the properties of the mass distribution that gives rise to it. 

In general relativity, the spacetimes outside bodies can similarly be described by a set of multipole moments. Spacetimes that are stationary, axisymmetric, reflection symmetric across the equatorial plane, and asymptotically flat -- an example being the Kerr black hole\index{Kerr black hole} -- are characterized by \emph{two} sets of multipole moments:\index{multipole moments} mass multipole moments $M_0$, $M_2$, $M_4$, ..., and current (or spin) multipole moments $S_1$, $S_3$, $S_5$, ... . $M_0 = M$ is the mass, $S_1 = J$ is the spin, and $M_2$ is the mass quadrupole moment. Now, according to the no hair theorem\index{no hair theorem} \cite{Hansen1974}, the multipole moments of quiescent black holes with the above properties satisfy
\begin{equation}
M_l + i S_l = M (i a)^l,
\label{multipoles}
\end{equation}
where $a = J/M$. Hence only two of them are independent: a quiescent black hole can be characterized completely by its mass $M$ and spin $J$.  Measuring any three of the multipole moments\index{multipole moments} and checking consistency with the above relation would constitute a test of general relativity. The $M_l$ and $S_l$ have dimensions of $\mbox{(mass)}^{l+1}$, and it is convenient to instead use dimensionless quantities $\mathfrak{m}_l = M_l/M^{l+1}$ and $\mathfrak{s}_l = S_l/M^{l+1}$, as we shall do below.

\subsection{Ringdown}

At the end of inspiral,\index{inspiral} binary neutron stars\index{binary neutron stars} or black holes plunge towards each other to form a single, highly excited black hole, which will then undergo ``ringdown"\index{ringdown} as it evolves to a quiescent, Kerr black hole.\index{Kerr black hole} This process can be modeled as perturbations around a Kerr background, subject to the Einstein equations. For a black hole with mass $M$ at a distance $D$, the ``plus" and ``cross" polarizations then take the form of damped sinusoids, the quasi-normal modes (QNMs) \cite{Vishveshwara1970}:\footnote{Strictly speaking there is another integer $n$ characterizing the polarizations \cite{Berti2006}. However, overtones with $n \neq 0$ have small amplitudes and short damping times; we will not consider them here.}
\begin{eqnarray}
h_+(t) &=& \frac{M}{D}  \sum_{l, m} A_{lm}\,Y_+^{lm}\,e^{-t/\tau_{lm}}\,\cos(\omega_{lm} t - m \phi), \label{hplusQNM} \\
h_\times(t) &=& - \frac{M}{D}  \sum_{l, m} A_{lm}\,Y_\times^{lm}\,e^{-t/\tau_{lm}}\,\sin(\omega_{lm} t - m \phi), \label{hcrossQNM} 
\end{eqnarray}
with $\phi$ a phase offset, and $Y_+^{lm}$, $Y_\times^{lm}$ are linear combinations of spin-weighted spherical harmonics\index{spin-weighted spherical harmonics} ${}_{-2}Y^{lm}$ \cite{Berti2007},
\begin{eqnarray}
Y_+^{lm}(\iota) &=& {}_{-2}Y^{lm}(\iota,0) + (-1)^l {}_{-2}Y^{l-m} (\iota,0), \\
Y_\times^{lm}(\iota) &=& {}_{-2}Y^{lm}(\iota,0) - (-1)^l {}_{-2}Y^{l-m} (\iota,0),
\end{eqnarray}
with $\iota$ the angle between the direction of the black hole's intrinsic angular momentum and the line of sight to the observer. 

The damping times and mode frequencies $\tau_{lm}(M, J)$, $\omega_{lm}(M, J)$ in Eqns.~(\ref{hplusQNM}), (\ref{hcrossQNM}) only depend on the black hole mass $M$ and its spin $J$ \cite{Ruffini1971}. Hence, in general relativity only two of the $\tau_{lm}$ and $\omega_{lm}$ are independent, which opens up the possibility of a test of GR, similar to the one described above for the case of inspiral.\index{inspiral} This would effectively be a test of the no hair theorem. Indeed, the reason why frequencies and damping times only depend on these two quantities is that (a) the background spacetime around which one considers perturbations is assumed to be Kerr, and (b) the perturbative Einstein equations are assumed valid on this spacetime background, forcing relationships between damping frequencies and times. 

Gossan, Veitch, and Sathyaprakash \cite{Gossan2012} demonstrated how one can exploit the interdependences of the $\tau_{lm}$, $\omega_{lm}$ to test GR with Einstein Telescope\index{Einstein Telescope} as well as space-based detectors. For simplicity, they assumed that the spins of the progenitor objects were zero, in which case the amplitudes $A_{lm}$ in (\ref{hplusQNM}), (\ref{hcrossQNM}) only depend on the symmetric mass ratio $\eta = m_1 m_2/(m_1 + m_2)^2$. Using data from numerical simulations in \cite{Kamaretsos2012a}, they arrived at an analytic fit for the amplitudes $A_{21}$, $A_{22}$, $A_{33}$, and $A_{44}$ of the four most dominant modes as a function of $\eta$, and the damping times $\tau_{lm}^{\rm GR}(M, J)$ and QNM frequencies $\omega_{lm}^{\rm GR}(M, J)$ predicted by GR were modeled using simple analytic fits from Berti \emph{et al.} \cite{Berti2006}. It was then assumed that the true damping times $\tau_{lm}$ and $\omega_{lm}$ might deviate from the GR prediction by dimensionless relative shifts $\Delta\hat{\tau}_{lm}$ and $\Delta\hat{\omega}_{lm}$, respectively:
\begin{eqnarray}
\tau_{lm} &=& (1 + \Delta \hat{\tau}_{lm})\,\tau^{\rm GR}_{lm}(M, J), \label{taunonGR} \\
\omega_{lm} &=& (1 + \Delta \hat{\omega}_{lm})\,\omega^{\rm GR}_{lm}(M, J), \label{omeganonGR}
\end{eqnarray}
where in the case of GR, $\Delta\hat{\tau}_{lm} = \Delta\hat{\omega}_{lm} = 0$ for all $l$ and $m$. The full parameter space for the ``deviating" model $\mathcal{H}_{\rm dev}$ was then 
\begin{equation}
\{\Delta \hat{\tau}_{lm}, \Delta\hat{\omega}_{lm}, \vec{\theta}\},
\label{allparams}
\end{equation}
with $\vec{\theta} = \{ M, J, \ldots \}$ the parameters of the GR waveform. In practice, only a limited number of frequencies and damping times were allowed to be non-zero, leading to a parameter space
\begin{equation} 
\{ \Delta\hat{\omega}_{22}, \Delta\hat{\tau}_{22}, \Delta\hat{\omega}_{33}, \vec{\theta} \}.
\end{equation}

With the second-generation detectors, it is unlikely that much more than the 22 mode will be distinguishable. As shown by Kamaretsos \emph{et al.} \cite{Kamaretsos2012a}, the situation is quite different for Einstein Telescope\index{Einstein Telescope} or a space-based detector such as LISA,\index{LISA} where the 21, 22, 33, and 44 modes can all contribute signicantly to the signal-to-noise ratio. Numerical experiments were performed in which simulated signals were added to stationary, Gaussian noise following the projected noise power spectral densities of Einstein Telescope and LISA. The sensitivity to GR violations of the type (\ref{taunonGR}), (\ref{omeganonGR}) was then checked by two methods:
\begin{itemize}
\item \emph{Direct parameter estimation.} Given data $d$ and the signal model $\mathcal{H}_{\rm dev}$, the posterior probability\index{posterior probability} distribution for the parameters $\vec{\lambda}$ of Eq.~(\ref{allparams}) is given by
\begin{equation}
p(\vec{\lambda} | d, \mathcal{H}_{\rm dev}, I) = \frac{p(d | \mathcal{H}_{\rm dev}, \vec{\lambda}, I)\,p(\vec{\lambda} | \mathcal{H}_{\rm dev}, I)}{p(d | \mathcal{H}_{\rm dev}, I)},
\label{PDF}
\end{equation}
where we have used Bayes' theorem.\index{Bayes' theorem} One has
\begin{equation}
p(d | \mathcal{H}_{\rm dev}, I) = \int d\vec{\lambda}\,p(d|\mathcal{H}_{\rm dev}, \vec{\lambda}, I)\,p(\vec{\lambda}|\mathcal{H}_{\rm dev}, I)
\end{equation}
with $p(\vec{\lambda} | \mathcal{H}_{\rm dev}, I)$ the prior distribution of the parameters. $p(d|\mathcal{H}_{\rm dev}, \vec{\lambda}, I)$ is the likelihood\index{likelihood function} of the data given parameters $\vec{\lambda}$, as in Eq.~(\ref{likelihoodfunction}):
\begin{equation}
p(d|\mathcal{H}_{\rm dev}, \vec{\lambda}, I) = \mathcal{N}\,e^{-( d - h(\vec{\lambda}) | d - h(\vec{\lambda})  )/2},
\end{equation}
where $h(\vec{\lambda}; t)$ is the waveform family corresponding to $\mathcal{H}_{\rm dev}$. Posterior distributions for parameters like $\Delta\hat{\tau}_{lm}$ and $\Delta\hat{\omega}_{lm}$ are obtained by integrating the posterior probability\index{posterior probability} density (\ref{PDF}) over all the other parameters. 
\item \emph{Model selection.} Here two models were considered: the GR model  $\mathcal{H}_{\rm GR}$ in which $\Delta\hat{\tau}_{lm} = \Delta\hat{\omega}_{lm} = 0$ and only the $\vec{\theta}$ are free parameters, and the ``deviating" model $\mathcal{H}_{\rm dev}$ in which the $\Delta\hat{\tau}_{lm}$ and $\Delta\hat{\omega}_{lm}$ are allowed to vary on top of the $\vec{\theta}$. One then computes an odds ratio
\begin{eqnarray}
O^{\rm dev}_{\rm GR} &=& \frac{P(\mathcal{H}_{\rm dev} |d, I)}{P(\mathcal{H}_{\rm GR} | d, I)} \\
               				   &=& \frac{P(\mathcal{H}_{\rm dev} | I)}{P(\mathcal{H}_{\rm GR} | I )} \frac{P(d | \mathcal{H}_{\rm dev}, I)}{P(d | \mathcal{H}_{\rm GR}, I)}.
\label{oddsringdown}
\end{eqnarray}
The ratio of prior probabilities, $P(\mathcal{H}_{\rm dev} | I) / P(\mathcal{H}_{\rm GR} | I )$, is just a constant overall prefactor, which for convenience can be set to one.
\end{itemize}

With parameter estimation, it was found that a 10\% deviation in $\omega_{22}$ would be clearly visible for a $500\,M_\odot$ black hole at 1.25 Gpc in ET, and for $10^6\,M_\odot$ and $10^8\,M_\odot$ at 1.25 Gpc and 10 Gpc, respectively, in LISA.\index{LISA} $500\,M_\odot$ coalescences are expected to be rare within distances of 1.25 Gpc. By contrast, the quoted mass and distance range for LISA is expected to correspond to a detection rate of tens per year \cite{Taskforce}. 

Bayesian model selection leads to rather better results. A 10\% deviation in $\omega_{22}$ would be visible for $500\,M_\odot$ at $D_{\rm L} \simeq 6$ Gpc in Einstein Telescope,\index{Einstein Telescope} and for $10^6\,M_\odot$ at a similar distance with LISA. A 0.6\% deviation could be picked up at a redshift of 5 with a $10^8\,M_\odot$ source in LISA. 

The above results are for black holes resulting from a binary with non-spinning components. However, in \cite{Kamaretsos2012b}, it was shown that in the case where both components have spins, the ringdown mode amplitudes $A_{lm}$ retain a memory not only of the progenitor's mass ratio, but also of spins. 

We end this subsection with an important comment. In the odds ratio (\ref{oddsringdown}), \emph{the hypothesis $\mathcal{H}_{\rm dev}$ is not the equivalent of our $\mathcal{H}_{\rm modGR}$ in the case of inspiral}\index{inspiral} (Sec.~\ref{subsec:TIGER}). To see this, denote the ``testing parameters" by
\begin{equation}
(\phi_1, \phi_2, \phi_3) \equiv (\Delta\hat{\omega}_{22}, \Delta\hat{\tau}_{22}, \Delta\hat{\omega}_{33}),
\end{equation}
where in this case, $\phi_i =0$ for $i = 1,2,3$ corresponds to GR. In the language of TIGER,\index{TIGER method} one then has $\mathcal{H}_{\rm dev} = H_{123}$, the hypothesis in which all three parameters \emph{at the same time} are different from the GR prediction. Indeed, in the ``deviating" waveform model, all of the testing parameters are allowed to vary freely, but \emph{e.g.}~each of the hypersurfaces $\phi_i = 0$ have zero measure in the model's parameter space, and zero prior probability.\index{prior probability} In this notation and with the prior odds for $\mathcal{H}_{\rm dev}$ against $\mathcal{H}_{\rm GR}$ set to some arbitrary $\alpha$,
\begin{equation}
O^{\rm dev}_{\rm GR} = \alpha B^{123}_{\rm GR}. 
\end{equation}
The resolvability of anomalies in the testing parameters would no doubt improve if instead one were to compute an odds ratio
\begin{eqnarray}
O^{\rm modGR}_{\rm GR} &=& \frac{P(\mathcal{H}_{\rm modGR}|d, I)}{P(\mathcal{H}_{\rm GR}|d, I)} \\
                                        &=& \frac{\alpha}{2^3-1} \sum_{k = 1}^{3}\sum_{i_1 < \ldots < i_k} B^{i_1 \ldots i_k}_{\rm GR},
\end{eqnarray}
completely analogously to (\ref{OddsSingleSource}), possibly with a larger number of testing parameters than just $\{\Delta\hat{\omega}_{22}, \Delta\hat{\tau}_{22}, \Delta\hat{\omega}_{33}\}$. It would also be of great interest to see what happens if information from multiple sources is combined,
\begin{eqnarray}
\mathcal{O}^{\rm modGR}_{\rm GR} = \frac{P(\mathcal{H}_{\rm modGR}|d_1 , \ldots, d_{\mathcal{N}}, I)}{P(\mathcal{H}_{\rm GR}|d_1 , \ldots, d_{\mathcal{N}}, I)},
\end{eqnarray}
analogously to (\ref{OddsCatalog}).

\subsection{Extreme mass ratio inspirals}

Extreme mass ratio inspirals\index{extreme mass ratio inspirals} (EMRIs) consist of a very massive black hole (or boson star \cite{Liebling2012}, or naked singularity, ...) surrounded by a smaller object, which could be a neutron star or a stellar mass black hole. In the case of Einstein Telescope,\index{Einstein Telescope} target systems would have a massive component of a few hundred solar masses \cite{Amaro-Seoane2010}, while for space-based detectors the mass would be in the range $10^5 - 10^9\,M_\odot$ \cite{Amaro-Seoane2012}. EMRIs provide another avenue to testing the no hair theorem: the orbits are expected to be extremely complicated, and in the case of LISA,\index{LISA} there will be a large number of gravitational wave cycles within the detector's frequency band. As a consequence, the gravitational wave emission of the smaller object will bear a detailed imprint of the spacetime in the vicinity of the massive object.

Ryan was the first to evaluate the measurability of multipole moments\index{multipole moments} using EMRI signals in Advanced LIGO\index{Advanced LIGO} and LISA \cite{Ryan1997}. With a number of simplifying assumptions -- the most important one being circularity of the orbit of the smaller object, which moreover is taken to move in the equatorial plane -- one can write down an expression for the phase in the Fourier domain explicitly showing the dependence on the multipoles  $\mathfrak{m}_l$, $\mathfrak{s}_l$. In particular, $\mathfrak{s}_1$ first appears at 1.5PN order, and $\mathfrak{m}_2$ at 2PN. In the case of Advanced LIGO, Ryan's conclusion was that even assuming $m_1 = 30\,M_\odot$ (a very heavy stellar mass black hole) and $m_2 = 0.2$ (an implausibly light neutron star), it would be hard to independently measure $\mathfrak{s}_1$ and $\mathfrak{m}_2$ with a single inspiral\index{inspiral} event near SNR threshold. For LISA, the situation is quite different. Assuming $m_1 = 10^5\,M_\odot$, $m_2 = 10\,M_\odot$, and SNR = 100, one obtains 1-$\sigma$ measurement accuracies of $\Delta \mathfrak{s}_1 \sim 10^{-4}$ and $\Delta \mathfrak{m}_2 \sim 1.5 \times 10^{-3}$. This would allow for a precision test of the no hair theorem. 

Subsequent to Ryan's seminal paper, a number of authors have relaxed his assumptions. Collins and Hughes developed a treatment of multipole moments\index{multipole moments} that is more appropriate than Ryan's in the strong-field regime \cite{Collins2004} through the notion of ``bumpy black  holes".\index{bumpy black holes} The motion of the smaller object will not be quasi-circular; Glampedakis and Babak employed so-called kludge waveforms which encode the essentials of the orbital motion \cite{Glampedakis2006}. Barack and Cutler \cite{Barack2007} showed that with kludge waveforms, results for the measurability of low-order multipole moments are qualitatively in keeping with those of Ryan. Vigeland and Hughes studied orbits around ``bumpy black holes" , showing how a spacetime's bumps are imprinted onto the orbital frequencies \cite{Vigeland2010}. Recently, Vigeland, Yunes and Stein studied bumpy black holes in alternative theories of gravity\index{alternative theories of gravity} \cite{Vigeland2011}.

Most recently, Rodriguez \emph{et al.} performed a more in-depth study of the possibility of using second-generation detectors to test the no hair theorem, also assuming a more reasonable mass for the lighter object ($1.4\,M_\odot$), and masses between $10\,M_\odot$ and $150\,M_\odot$ for the heavier one \cite{Rodriguez2012}. This work still mostly considered parameter estimation; it would be of great interest to cast the problem in terms of hypothesis testing, in which case results from multiple sources could be combined.

\section{Probing the large-scale structure of spacetime}
\label{sec:cosmography}

\subsection{Binary inspirals as standard sirens}

Assuming that at large scales the Universe is homogeneous and isotropic, its line element can be put in the Friedman-Lema\^{i}tre-Robertson-Walker (FLRW) form \cite{MTW}:
\begin{equation}
ds^2 = - dt^2 + a^2(t)\,\left[\frac{dr^2}{1-kr^2} + r^2\left( d\theta^2 + \sin^2\theta\,d\phi^2 \right)\right],
\end{equation}
where the entire dynamical content resides in the evolution of the scale factor $a(t)$. The constant $k$ can be positive, zero, or negative, in which case the $t = const$ spatial hypersurfaces are hyperspherical, flat, or hyperboloidal, respectively. Given a homogeneous mass distribution $\rho$ with pressure $P$, the Einstein equations reduce to two equations for $a(t)$, $\rho(t)$, and $P(t)$, called the Friedman equations,\index{Friedman equations}
\begin{eqnarray}
\left(\frac{\dot{a}}{a}\right)^2 &=& \frac{8\pi}{3} \rho -\frac{k}{a^2}, \label{Friedman1} \\
\frac{\ddot{a}}{a}     			  &=& - \frac{4\pi}{3} \left( \rho + 3 P \right), \label{Friedman2}
\end{eqnarray}
which can be combined to arrive at an equation for the time evolution of the density:
\begin{equation}
\dot{\rho} = -3\,(\rho + P)\,\frac{\dot{a}}{a}.
\label{rhodot}
\end{equation}
This can be solved given an equation of state $P = P(\rho)$. In the case of pressureless dust (which can serve as a model for a sprinkling of galaxies), $P = 0$, and $\rho \propto a^{-3}$. For radiation, $P = \rho/3$, leading to $\rho \propto a^{-4}$. Finally, there is evidence that the expansion of the Universe is speeding up \cite{Riess1998}. The cause is unknown, but it is convenient to model it as \emph{dark energy},\index{dark energy} a perfect fluid with positive density but negative pressure. Postulating an equation of state of the form $P(t) = w(t)\,\rho(t)$ with $w(t) < 0$, one can once again solve Eq.~(\ref{rhodot}) to obtain an expression for $\rho$ as a function of the scale factor $a$.

Using this and the first Friedman equation (\ref{Friedman1}), the way the Universe evolves depending on its contents can be expressed through the \emph{Hubble parameter}\index{Hubble parameter} $H(a)$, defined as
\begin{eqnarray}
H^2(a) &\equiv& \left(\frac{\dot{a}}{a}\right)^2 \\
           &=& H_0^2 \left[ \Omega_{\rm M} a^{-3} + \Omega_{\rm R} a^{-4} + \Omega_k a^{-2} + \Omega_{\rm DE} \exp\left( 3 \int_0^a \frac{da'}{a'}\,\left[ 1 + w(a') \right] \right) \right], \nonumber\\
\end{eqnarray}
where $H_0$ is the Hubble constant,\index{Hubble constant} which gives the expansion of the Universe\index{expansion of the Universe} at the current epoch, and the dimensionless quantities $\Omega_{\rm M}$, $\Omega_{\rm R}$, $\Omega_k$, and $\Omega_{\rm DE}$ are the fractional contributions to the total density of, respectively, matter, radiation, spatial curvature,\index{spatial curvature} and dark energy:\index{dark energy}
\begin{equation}
\Omega_{\rm M} = \frac{8\pi}{3 H_0^2}\rho_{M,0}, \,\,\,\,\,\,\,\Omega_{\rm R} = \frac{8\pi}{3 H_0^2}\rho_{\rm R,0},\,\,\,\,\,\,\,\Omega_k = - \frac{k}{H_0^2},\,\,\,\,\,\,\,\Omega_{\rm DE} = \frac{8\pi}{3 H_0^2}\rho_{\rm DE,0},
\end{equation}
with $\rho_{\rm M,0}$, $\rho_{\rm R,0}$, $\rho_{\rm DE,0}$ the densities at the current epoch of matter, radiation, and dark energy,\index{dark energy} respectively. 

An important task in cosmology is to determine 
\begin{equation}
\vec{\Omega} \equiv (H_0, \Omega_{\rm M}, \Omega_{\rm R}, \Omega_k, \Omega_{\rm DE}, w(t)), 
\label{cosmoparams}
\end{equation}
and especially to gain empirical insight into the enigmatic dark energy, by measuring its equation-of-state parameter $w(t)$. The main tools for studying the late-time evolution of the Universe are \emph{standard candles}.\index{standard candles} These are distance markers for which both the redshift $z$\index{redshift} and the luminosity distance\index{luminosity distance} $D_{\rm L}$ are known. For a source with intrinsic luminosity $\mathcal{L}$ and observed flux $\mathcal{F}$, $D_{\rm L}$ is defined through
\begin{equation}
\mathcal{F} = \frac{\mathcal{L}}{4\pi D_{\rm L}^2}.
\end{equation}
If the Universe were Euclidean and never-changing, $D_{\rm L}$ would correspond to the familiar Euclidean notion of distance. However, due to the evolution of the Universe, $D_{\rm L}$ and $z$ are related in a complicated way:
\begin{equation}
D_{\rm L}(z) 
= c\,(1+z)\,
\left\{
\begin{array}{ll}
|k|^{-1/2} \sin\left[ |k|^{1/2} \int_0^z \frac{dz'}{H(z')} \right] & \mbox{for}\,\,\, \Omega_k < 0, \\
\int_0^z \frac{dz'}{H(z')} 									      & \mbox{for}\,\,\,\Omega_k = 0, \\
|k|^{-1/2} \sinh\left[ |k|^{1/2} \int_0^z \frac{dz'}{H(z')} \right] & \mbox{for}\,\,\, \Omega_k > 0, 
\end{array}
\right.
\label{DL}
\end{equation}
where $H(z)$ is the Hubble parameter\index{Hubble parameter} as a function of redshift. Since radiation will not be very important at late times, one can write (using $1/a  = 1 + z$)
\begin{equation}
H(z) = H_0\,\left[\Omega_{\rm M}(1 + z)^3 + \Omega_k (1+z)^2 + (1 - \Omega_{\rm M} - \Omega_k)\,E(z)\right]^{1/2},
\label{Hubbleparameter}
\end{equation} 
where $E(z)$ depends on the equation of state of dark energy.\index{dark energy} At late times, one can expand $w(t)$, or equivalently $w(z)$, as
\begin{eqnarray}
w(z) &=& P_{\rm DE}/\rho_{\rm DE} = w_0 + w_a (1-a) + \mathcal{O}\left[(1-a)^2\right] \\
       &\simeq& w_0 + w_a \frac{z}{1+z},
\label{wexpansion}
\end{eqnarray}
in which case
\begin{equation}
E(z) = (1 + z)^{3(1 + w_0 + w_a)}\,e^{-3w_a z/(1+z)}.
\end{equation}
From Eqns.~(\ref{DL}) and (\ref{Hubbleparameter}), it will be clear that given a large number of astrophysical sources for which the pairs $(D_{\rm L}, z)$ can be measured, one can constrain the parameters (\ref{cosmoparams}).

The most commonly used standard candles\index{standard candles} are Type Ia supernovae,\index{Type Ia supernovae} whose luminosity is believed to be known within $\sim 10\%$ \cite{Riess1998}. However, this luminosity needs to be calibrated by comparison with different kinds of closer-by sources, leading to a ``cosmic distance ladder",\index{cosmic distance ladder} each rung of which could contain unknown systematic errors. As pointed out by Schutz in 1986, GW signals from inspiraling neutron stars and black holes can provide an absolute measure of distance, with no dependence on other sources \cite{Schutz1986}. In the context of cosmology, they have been dubbed \emph{standard sirens}.\index{standard sirens} To see how this works, consider the amplitude of an inspiral\index{inspiral} signal as a function of time:
\begin{equation}
\mathcal{A}(t) = \frac{1}{D_{\rm L}} \mathcal{M}^{5/3} g(\theta, \phi, \iota, \psi)\,\omega^{2/3}(t).
\label{GWamplitude}
\end{equation}
Here $\mathcal{M} = (m_1 m_2)^{3/5}/(m_1 + m_2)^{1/5}$ is the chirp mass,\index{chirp mass} $g(\theta, \phi, \iota, \psi)$ is a known function of the sky position $(\theta,\phi)$ and orientation of the orbital plane $(\iota,\psi)$, and $\omega(t) = \dot{\Phi}(t)$ is the instantaneous frequency. The chirp mass, and of course $\omega(t)$, can be obtained from the phase. Thus, if sky position and orientation are known, then from the amplitude one can infer the luminosity distance $D_{\rm L}$. 

\subsection{Cosmography with gravitational wave detectors}

To make use of binary inspirals\index{inspiral} as standard sirens, what is needed is a way to obtain some information about redshift $z$, and also about the sky position $(\theta,\phi)$ and orientation $(\iota,\psi)$ so that the luminosity distance $D_{\rm L}$ can be obtained from the GW amplitude. A variety of methods have been proposed to achieve this. 

\subsubsection{Using electromagnetic counterparts}

Gamma ray bursts\index{gamma ray bursts} (GRBs) are among the most energetic electromagnetic events since the Big Bang. They roughly fall into two categories: short, hard GRBs and long, soft ones. It is believed (although only direct GW detection will provide a definitive answer) that short, hard GRBs are caused by the coalescence of two neutron stars, or a neutron star and a black hole \cite{Nakar2007}. Sometimes a GRB can be localized on the sky, providing $(\theta,\phi)$. If this allows for the identification of the galaxy that was host to the inspiral\index{inspiral} event, then from its spectrum one can infer the redshift $z$. Finally, given a \emph{network} of detectors, some information about the orientation $(\iota,\phi)$ can also be obtained. Additionally, it is possible that GRBs are strongly beamed in a direction perpendicular to the inspiral plane, with inclination angle $\iota \lesssim 20^\circ$.

In \cite{Nissanke2010}, Nissanke \emph{et al.} investigated with what accuracy a network of advanced detectors would be able to do cosmology. We first note that with second-generation detectors, the maximum redshift out to which inspirals can be seen is $z \simeq 0.1$ for BNS and $z \simeq 0.2$ for NSBH. For small redshifts, the luminosity distance-redshift relationship, Eq.~(\ref{DL}), reduces to
\begin{equation}
D_{\rm L} \simeq \frac{cz}{H_0},
\label{HubbleLaw}
\end{equation}
which is just Hubble's law.\index{Hubble's law} This means that with advanced detectors, we will only be able to probe the Hubble constant\index{Hubble constant} $H_0$. However, since $H_0$ is an overall scaling factor in the full expression for $D_{\rm L}$, its accurate and unbiased measurement is key to precision cosmology at the largest scales. Gravitational wave detection will provide us with a way of measuring $H_0$ without having to rely on any kind of cosmic distance ladder.\index{cosmic distance ladder} Note that from Eq.~(\ref{HubbleLaw}), if redshift can be determined with essentially zero uncertainty, then the uncertainty $\Delta H_0$ on the Hubble parameter is related to the distance uncertainty by
\begin{equation}
\frac{\Delta H_0}{H_0} = \frac{\Delta D_{\rm L}}{D_{\rm L}}.
\end{equation}
This pertains to a \emph{single} source; the accuracy will improve roughly as $\sim \sqrt{\mathcal{N}}$ for $\mathcal{N}$ events. Nissanke \emph{et al.} found that with a network composed of the two Advanced LIGOs\index{Advanced LIGO} and Advanced Virgo,\index{Advanced Virgo} with $\mathcal{N} = 4$ BNS events one already has $\Delta H_0/H_0 \sim 13\%$, and with $\mathcal{N} = 15$, $\Delta H_0/H_0 \sim 5\%$ \cite{Nissanke2010}. 

\begin{figure}[htbp!]
\centering
\includegraphics[height=6cm]{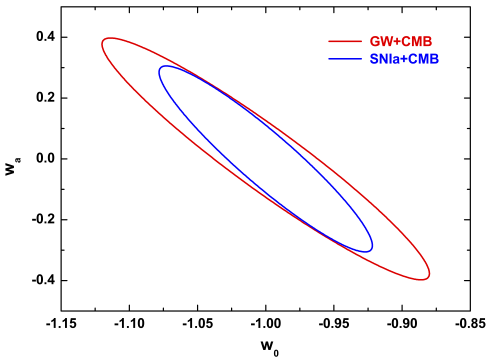}
\caption{Measurement uncertainties for the possible time dependence in the dark energy\index{dark energy} equation of state parameter $w$, modeled as $w(a) = w_0 + (1 - a) w_a$, with $a$ the scale factor \cite{Zhao2011}. The slightly larger, red error ellipse is for standard sirens as seen with ET, the blue one for the possible future SNAP Type Ia supernova\index{Type Ia supernovae} survey \cite{SNAP2004}. In both cases, prior information from the CMB is assumed for $\Omega_{\rm M}$, $\Omega_{\rm DE}$, and $\Omega_k$.}
\label{fig:w0wa}
\end{figure}

With Einstein Telescope\index{Einstein Telescope} it would be possible to see BNS events out to redshifts of several. In \cite{Sathyaprakash2010} and \cite{Zhao2011}, detailed studies were made of how accurately the \emph{full} set of cosmological parameters (\ref{cosmoparams}) could be measured. With $\mathcal{O}(1000)$ events with identifiable host galaxies over the course of 5-10 years, $\Omega_{\rm M}$ and $\Omega_{\rm DE}$ could be constrained with an uncertainties comparable to what one finds in measurements of the Cosmic Microwave Background\index{Cosmic Microwave Background} (CMB). One can also use the CMB measurements for $\Omega_{\rm M}$, $\Omega_{\rm DE}$, and $\Omega_k$, and their uncertainties, as priors, and focus on the dark energy\index{dark energy} equation of state parameter $w$, including its possible time dependence. Using a linear approximation to $w(z)$ as in Eq.~(\ref{wexpansion}), one can then compare accuracies in measuring $w_0$ and $w_a$, on the one hand using standard sirens seen by ET, and on the other hand considering the SNAP Type Ia supernova\index{Type Ia supernovae} survey which may be available on the same timescale as ET \cite{SNAP2004}. The results are shown in Fig.~\ref{fig:w0wa}. The measurement quality is comparable in the two cases, but we stress once again that standard sirens allow for an \emph{independent} measurement, with no need for a cosmic distance ladder.\index{cosmic distance ladder} 

\subsubsection{Using a prior on the intrinsic neutron star masses}

Currently there are about 10 electromagnetically observed binary neutron star systems, with varying degrees of compactness. The distribution of neutron star masses in these binaries is relatively tight \cite{Kiziltan2010}, with mean $\mu_{\rm NS} \simeq 1.34\,M_\odot$ and standard deviation $\sigma_{\rm NS} \simeq 0.06\,M_\odot$. Now, the masses that are measured from a gravitational wave signal are not the physical ones $m_{\rm phys}$, but the redshifted masses $m_{\rm obs}$; for a source at redshift $z$, one has 
\begin{equation}
m_{\rm obs} = (1 + z)\,m_{\rm phys}.
\end{equation}
As shown by Taylor, Gair, and Mandel, assuming an underlying, physical distribution of masses and comparing with the observed masses, one can obtain information about the redshifts of events without ever needing an electromagnetic counterpart \cite{Taylor2012a}. With $\sim 100$ BNS observations, a network of second-generation detectors would then allow the measurement of the Hubble constant\index{Hubble constant} with $\sim 10\%$ uncertainty. 

Taylor and Gair applied this idea to a network of Einstein Telescopes,\index{Einstein Telescope} in which case one might have as many as $10^5$ BNS signals per year \cite{Taylor2012b}. Keeping $H_0$, $\Omega_{\rm M}$, $\Omega_{\rm DE}$, and $\Omega_k$ fixed, they found that also with this method, the dark energy\index{dark energy} equation of state parameters $(w_0, w_a)$ in Eq.~(\ref{wexpansion}) could be measured with an accuracy comparable to the forecasted constraints from future SNIa surveys with CMB and other results as priors. 

\subsubsection{Using galaxy catalogs}

\begin{figure}[htbp!]
\centering
\includegraphics[height=6cm]{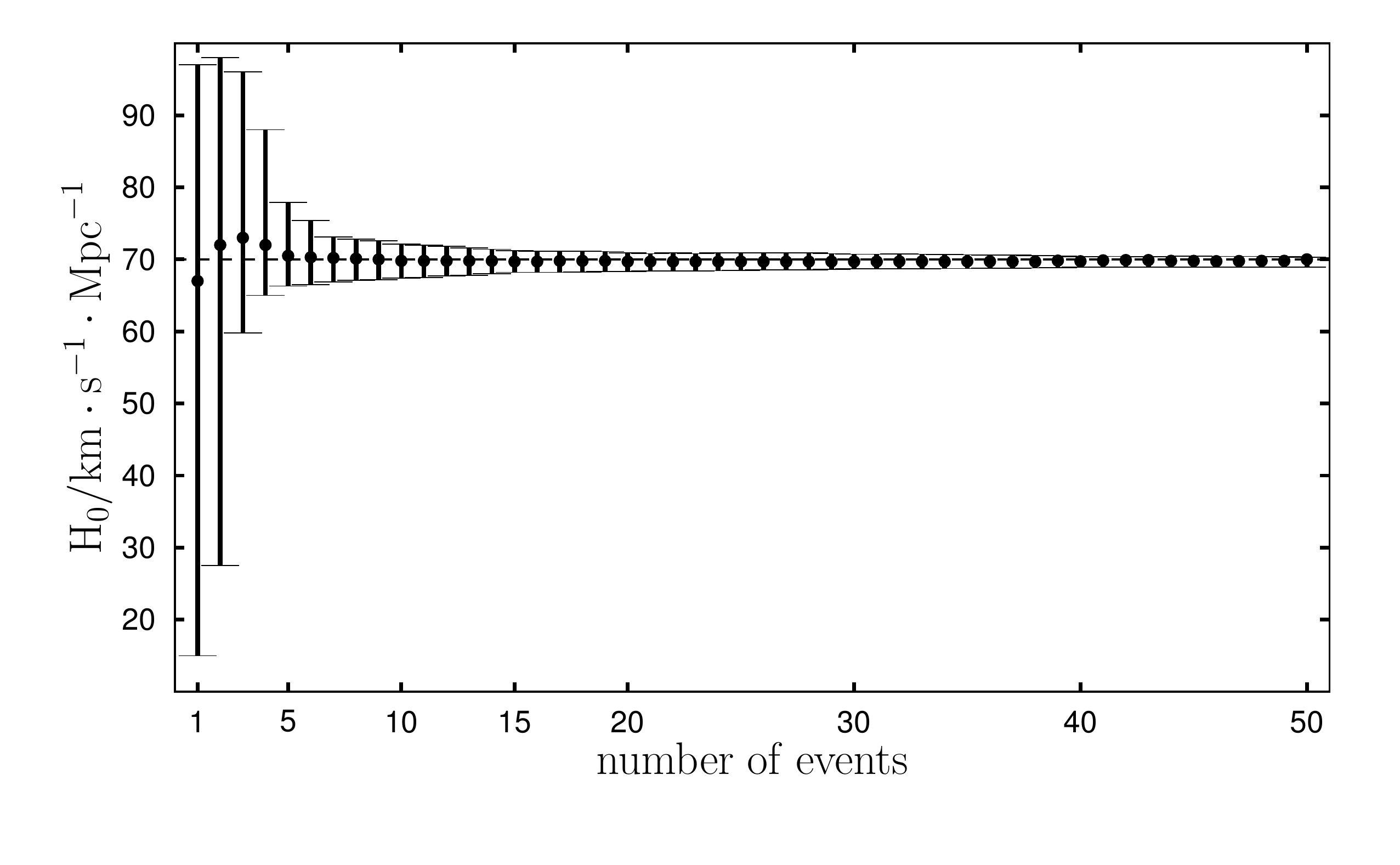}
\caption{Evolution of the medians and 95\% confidence intervals for the Hubble constant as information from an increasing number of coalescence events is combined, using a galaxy catalog to obtain information on redshifts, for the HLVJI network of second-generation detectors \cite{DelPozzo2012}.}
\label{fig:DelPozzo}
\end{figure}

Another exciting idea for measuring $H_0$ without the need for electromagnetic counterparts, and without having to restrict to a particular kind of inspiral\index{inspiral} event, was recently put forward by Del Pozzo \cite{DelPozzo2012}. He assumed three kinds of networks: the Advanced LIGO\index{Advanced LIGO} detectors together with Advanced Virgo\index{Advanced Virgo} (HLV), the same with the Japanese KAGRA\index{KAGRA} added (HLVJ), and the five-detector network with IndIGO\index{IndIGO} also included (HLVJI). Given an inspiral event, these networks will be able to localize it on the sky with different uncertainties; similarly, the distance $\hat{D}_{\rm L}$ extracted from the gravitational wave signal will also be subject to errors. Combining these uncertainties, one obtains a large 3-dimensional box in which the inspiral event could have occurred. A galaxy catalog\index{galaxy catalog} will yield a list of potential host galaxies within this box, all having different redshifts $\hat{z}_i$. Using the Hubble law\index{Hubble's law} $D_{\rm L} = cz/H_0$, the maximum-likelihood distance $\hat{D}_{\rm L}$, together with this list of redshifts, leads to a list of possible values for the Hubble constant,\index{Hubble constant} $\{ H_{0,i} \}$. However, a \emph{second} inspiral event will yield another list $\{ H_{0,j} \}$, which will typically have only limited overlap with the first one. As more and more detections are made, the true value of the Hubble constant will quickly emerge. Del Pozzo cast this idea into the language of Bayesian analysis, and found that \emph{after only 10 observations}, the 95\% confidence level accuracy on $H_0$ is 14.5\%, 7\%, and 6.7\% for the HLV, HLVJ, and HLVJI networks, respectively. After 50 observations, these numbers become $5\%$, $2\%$ and $1.8\%$, respectively; see also Fig.~\ref{fig:DelPozzo}. 

The advantage of this method is that it does not rely on any specific kind of source -- in principle it can use \emph{all} BNS, NSBH, and BBH detections. One issue will be the completeness of galaxy catalogs. However, it is possible to include in the analysis terms that describe the likelihood of observing a GW whose host galaxy was not detected by any survey because of its faintness; see \cite{Messenger2012a}.

MacLeod and Hogan explored a similar idea for measuring $H_0$ in the context of LISA,\index{LISA} with GW signals from EMRIs, and using galaxy clustering\index{galaxy clustering} \cite{MacLeod2008}. Finally, a supermassive binary black hole\index{supermassive binary black hole} at $z \lesssim 1$ would be sufficiently localizable with LISA that one might be able to find the host galaxy cluster, which would then yield a redshift, allowing for a measurement of the equation-of-state parameter of dark energy\index{dark energy} to within a few percent \cite{Iyer2007}.

\subsubsection{Using the neutron star equation of state to extract redshift from the gravitational wave signal}

Recently, a method was developed to extract redshift information from the GW signal itself, at least in the case of BNS or NSBH. In the last stages of inspiral,\index{inspiral} a neutron star will be deformed and acquire a quadrupole moment $Q_{ij}$ due to the tidal field $\mathcal{E}_{ij}$ of the companion object, and to leading order one can write $Q_{ij} = - \lambda \mathcal{E}_{ij}$. Here $\lambda$ is the tidal deformability parameter,\index{tidal deformability parameter} which depends both on the neutron star mass and the equation of state. These deformations have an influence on the orbital motion, which in turn gets imprinted onto the gravitational waveform. Such effects appear in the phase at 5PN and 6PN orders:
\begin{equation}
\Phi(v) = \Phi_{\rm PP}(v) + \Phi_{\rm tidal}(v),
\end{equation}
where $\Phi_{\rm PP}$ is the post-Newtonian phase under the assumption of point particles, and \cite{Hinderer2010,Pannarale2011}
\begin{eqnarray}
\Phi_{\rm tidal} &=& \sum_{a=2}^2 \frac{3\lambda_a}{128 \eta M^5}\left[ -\frac{24}{\chi_a} \left(1 + \frac{11 \eta}{\chi_a}\right)\,\left(\frac{v}{c}\right)^5 \right. \nonumber\\
&& \left. -\frac{5}{28 \chi_a} \left(3179 - 919\,\chi_a - 2286\,\chi_a^2 + 260\,\chi_a^3 \right)\,\left(\frac{v}{c}\right)^7
\right],
\label{Phitidal}
\end{eqnarray}
where the sum is over the components of the binary, $\chi_a = m_a/M$, and $\lambda_a = \lambda(m_a)$, $a = 1, 2$. The function $\lambda(m)$ takes the form $\lambda(m) = (2/3)\,k_2\,R^5(m)$, with $k_2$ the second Love number and $R(m)$ a neutron star's radius as a function of mass. Note that $\lambda(m)$ enters Eq.~(\ref{Phitidal}) only in the combination 
\begin{equation}
\frac{\lambda(m)}{M^5} \propto \left(\frac{R}{M}\right)^5 \sim 10^5.
\label{lambdaM5}
\end{equation}
Thus, although the tidal terms only appear at very high post-Newtonian order, they come with a large prefactor. Their effect will be noticeable already in advanced detectors, and certainly in Einstein Telescope.\index{Einstein Telescope}

Messenger and Read noted that the tidal contribution to the phase (\ref{Phitidal}) only depends on \emph{intrinsic} quantities \cite{Messenger2012b}. Indeed, the expansion of the Universe as the signal travels from source to observer will cause the ``observed" radius and mass to both be larger than the physical ones by a factor $(1+z)$, which however will cancel from (\ref{lambdaM5}) and hence (\ref{Phitidal}). 

Einstein Telescope might see $\mathcal{O}(10^5)$ BNS sources. Some fraction of these could be used to determine the neutron star equation of state by measuring $\lambda(m)$. Once this is done, for each source in the other fraction one would be able to determine the \emph{observed} masses $m_{\rm obs} = (1+z)\,m_{\rm phys}$ from the low-order PN  contributions to the phase, and the \emph{intrinsic} masses $m_{\rm phys}$ from the tidal contribution (\ref{Phitidal}). Hence both luminosity distances $D_{\rm L}$ and redshifts $z$ can be inferred directly from the gravitational wave signals! 

This will again allow for a fit of the luminosity distance--redshift relation $D_{\rm L}(z)$, thus constraining the cosmological parameters $\vec{\Omega}$ of Eq.~(\ref{cosmoparams}), on condition that the uncertainties on redshift measurements are not too high. The latter depend on the equation of state, about which not much is currently known. Messenger and Read estimate that in the range $z =  0.1 - 1$, $\Delta z/z \sim 0.1$ for the ``hardest" predicted equations of state (implying the greatest deformability), and $\Delta z/z \sim 0.4$ for ``soft" equations of state. How this translates into constraints on the parameters $\vec{\Omega}$ is yet to be investigated.

\section{Summary}
\label{sec:summary}

In a few years' time, the second-generation gravitational wave detectors are due to deliver their first detections. This will herald a new era in the empirical study of gravitation. For the first time, we will have access to the genuinely strong-field dynamics of gravity. As a bonus, we will be able to look at weak-field gravity in a novel way, by searching for effects that may only show up at very large distance scales, such as the ones which gravitational waves must travel from source to observer. 

In older literature, studies of how well one might test GR with gravitational waves mostly took the form of estimates. With the advanced detector era approaching, the last few years have seen the development of hands-on data analysis pipelines to look for deviations from GR in actual detector data. Soon we will be searching for alternative polarization states,\index{polarization states} as well as for possible anomalies in the way that dynamical gravitational fields interact with themselves. For the latter, a full data analysis pipeline using coalescences of binary neutron stars\index{binary neutron stars} is already in place. Binary black holes have a much richer dynamics, but the added complexity also makes for a formidable data analysis problem, the exploration of which has only just begun.

With third-generation ground-based detectors such as Einstein Telescope,\index{Einstein Telescope} and the space-based LISA,\index{LISA} one would be able to exploit not just the inspiral\index{inspiral} phase, but also the ringdown. As with the phasing coefficients, the Einstein equations impose relationships between characteristic frequencies and damping times, which effectively allow for a test of the no hair theorem. This will complement the test proposed earlier by Ryan, using extreme mass ratio inspirals.

Binary inspirals are ``standard sirens" which can be used to probe the large scale structure of spacetime. Although the basic idea had already been proposed by Schutz as early as 1986, the last few years have seen the development of detailed methods, based on electromagnetic counterparts, exploiting the mass distribution of binary neutron stars, using galaxy catalogs, or employing knowledge of the neutron star equation of state. 

The coming years are almost guaranteed to be a bonanza for gravitational physics. Either general relativity will be confirmed with more stringent tests than any that have been performed hitherto, or we will see deviations, which may well take the form of low-energy limits of quantum gravity effects. Any which way, the prospects are exciting indeed.

\section*{Acknowledgements}

The author is supported by the research programme of the Foundation for Fundamental Research on Matter (FOM), which is partially supported by the Netherlands Organisation for Scientific Research (NWO). It is a pleasure to thank M.~Agathos, K.G.~Arun, J.F.J.~van den Brand, N.~Cornish, W.~Del Pozzo, K.~Grover, M.~Hendry, I.S.~Heng, B.R.~Iyer, T.G.F.~Li, I.~Mandel, C.~Messenger, C.K.~Mishra, A.~Pai, M.~Pitkin, B.S.~Sathyaprakash, B.F.~Schutz, P.S.~Shawhan, T.~Sidery, R.~Sturani, M.~Tompitak, M.~Vallisneri, A.~Vecchio, J.~Veitch, S.~Vitale, and N.~Yunes,  for fruitful discussions. I would like to acknowledge the LIGO Data Grid clusters, without which some of the simulations described here could not have been performed. Specifically, these include the computing resources supported by National Science Foundation awards PHY-0923409 and PHY-0600953 to UW-Milwaukee. Also, I thank the Albert Einstein Institute in Hannover, supported by the Max-Planck-Gesellschaft, for use of the Atlas high-performance computing cluster.


\end{document}